\documentclass[12pt,epsf,amssymb]{article}
\usepackage{epsf,psfig}
\usepackage{graphicx}
\usepackage{amsfonts}
\usepackage[usenames]{color}
\usepackage{amsmath}

\begin{document}
\baselineskip 0.6cm
\newcommand{\gsim}{ \mathop{}_{\textstyle \sim}^{\textstyle >} }
\newcommand{\lsim}{ \mathop{}_{\textstyle \sim}^{\textstyle <} }
\newcommand{\vev}[1]{ \left\langle {#1} \right\rangle }
\newcommand{\bra}[1]{ \langle {#1} | }
\newcommand{\ket}[1]{ | {#1} \rangle }
\newcommand{\Dsl}{\mbox{\ooalign{\hfil/\hfil\crcr$D$}}}
\newcommand{\nequiv}{\mbox{\ooalign{\hfil/\hfil\crcr$\equiv$}}}
\newcommand{\nsupset}{\mbox{\ooalign{\hfil/\hfil\crcr$\supset$}}}
\newcommand{\nni}{\mbox{\ooalign{\hfil/\hfil\crcr$\ni$}}}
\newcommand{\EV}{ {\rm eV} }
\newcommand{\KEV}{ {\rm keV} }
\newcommand{\MEV}{ {\rm MeV} }
\newcommand{\GEV}{ {\rm GeV} }
\newcommand{\TEV}{ {\rm TeV} }
\newcommand{\bfx}{ {\mathbf x} }
\newcommand{\bfy}{ {\mathbf y} }
\newcommand{\bfk}{ {\mathbf k} }

\def\diag{\mathop{\rm diag}\nolimits}
\def\tr{\mathop{\rm tr}}

\def\Spin{\mathop{\rm Spin}}
\def\SO{\mathop{\rm SO}}
\def\O{\mathop{\rm O}}
\def\SU{\mathop{\rm SU}}
\def\U{\mathop{\rm U}}
\def\Sp{\mathop{\rm Sp}}
\def\SL{\mathop{\rm SL}}
\def\simgt{\mathrel{\lower2.5pt\vbox{\lineskip=0pt\baselineskip=0pt
           \hbox{$>$}\hbox{$\sim$}}}}
\def\simlt{\mathrel{\lower2.5pt\vbox{\lineskip=0pt\baselineskip=0pt
          \hbox{$<$}\hbox{$\sim$}}}}

\def\change#1#2{{\color{blue} #1}{\color{red} [#2]}\color{black}\hbox{}}


\begin{titlepage}

\begin{flushright}
UCB-PTH-06/21 \\
LBNL-62015 \\
\end{flushright}

\vskip 2cm
\begin{center}
{\large \bf Density Perturbations in Chain Inflation} 

\vskip 1.2cm
Brian Feldstein, Brock Tweedie

\vskip 0.4cm
{\it Department of Physics and Lawrence Berkeley National 
Laboratory,

University of California, Berkeley, CA 94720, USA} \\

\vskip 1.5cm

\abstract{We consider the model of ``Chain Inflation,'' in which the period of inflation 
in our universe took the form of a long sequence of quantum tunneling events.  We find that 
in the simplest such scenario, in which the tunneling processes are uniform, approximately 
$10^4$ vacua per e-folding of inflation are required in order that the density perturbations 
produced are of an acceptable size.  We arrive at this conclusion through a combination of 
analytic and numerical techniques, which could also serve as starting points for calculations 
with more general sets of assumptions.} 

\end{center}
\end{titlepage}



\section{Introduction}

In 1980, Guth proposed the inflationary universe scenario to solve the horizon and flatness 
problems \cite{Guth}.  In this picture, the very early universe experienced a period of 
exponential expansion driven by the vacuum energy of a scalar field.  In the original model, 
the scalar field was supposed to be stuck in a false vacuum of its potential.  After some time, 
however, quantum mechanical tunneling would bring the field to the true minimum of the potential 
and inflation would come to an end.

The tunneling process in quantum field theory was originally described by Coleman \cite{Coleman}.  
A tunneling event is characterized by the formation of a ``bubble,'' in which the inside of the 
bubble is located in the new true-vacuum phase, while the outside of the bubble is located in 
the old false-vacuum phase.  The vacuum energy lost in the interior of the bubble due to its 
formation is stored in the bubble wall;  energy is conserved in this process.  The bubble 
subsequently expands, at a speed quickly approaching the speed of light, eating up the false 
vacuum in favor of the true one.  When two bubbles collide the energy stored in the walls is 
converted into a spray of particles, which eventually thermalizes.

Unfortunately, Guth's original model had a fatal flaw:  If the tunneling rate from the false 
vacuum was too large, then there was too short a period of inflation to yield the 60 e-foldings 
of expansion necessary to solve the horizon and flatness problems.  If the tunneling rate was 
too small, then the true vacuum bubbles formed too far away from each other to ever have consistent 
collisions and percolation.  The interiors of the bubbles would then tend to remain forever empty, 
with the energy being stored permanently in the bubble walls.\footnote{Bubble collisions do occasionally 
happen in this scenario, but the inhomogeneities in the resulting radiation tend to be much too large.}  
It was argued in \cite{GW} that there is likely no intermediate value for the tunneling rate which 
can yield an acceptable number of e-foldings while also allowing bubbles to collide and percolate.

As a result of this problem, models of inflation based on slowly rolling fields \cite{slowroll1} 
\cite{slowroll2}, rather than vacuum tunneling, have become the standard paradigm.  On the other hand, 
Freese and Spolyar, in their paper ``Chain Inflation'' \cite{Freese1}, proposed a simple way to extend 
and save Guth's false vacuum style inflation.  Their idea was to choose a tunneling rate on the high 
side, so that bubbles would indeed percolate, but at the same time have a long sequence, or ``chain,'' 
of vacua to ensure an acceptable number of e-foldings.  In each step of the chain, only a fraction 
of an e-folding is obtained, but overall, inflation can be sustained for a sufficiently long time.  
The biggest open question in this model is whether or not it can lead to a phenomenologically 
acceptable spectrum of density perturbations.  The density perturbations need to be scale invariant, 
of an appropriate amplitude, and have very small departures from gaussianity.  It is not at all 
clear how such a spectrum could arise in chain inflation.

As noted in \cite{Freese1} the main source of density perturbations in chain inflation should come 
from the intrinsic randomness of the tunneling process itself.  Different points in space will reach 
the final step in the chain at different times, leading to inhomogeneities.  Scale invariance can 
most easily be satisfied by assuming a time translation symmetry amongst the tunneling events -- i.e., 
by assuming all steps in the chain are essentially  identical.  An example model with this feature 
is an axion with a tilted cosine potential, as discussed in \cite{Freese2}.  In this paper we will 
concentrate on this case, although we will have  comments concerning extensions to non-uniform tunneling 
as well.  Calculating the density perturbations analytically in chain inflation is extremely difficult, 
and we will instead use computer simulations for much of our analysis.

During the period of inflation responsible for the observed  modes of the density perturbations, the 
Hubble parameter $H$  may be taken as roughly constant.  Then the only free dimensionless parameter 
on which the amplitude of the perturbations can depend is $\Gamma^{-1} H^4$, where $\Gamma$ is the 
tunneling rate per unit volume for the uniform tunneling events.  The constraint of bubble percolation 
requires that $\Gamma^{-1} H^4 \lsim 1$, but since the amplitude of the perturbations is required to be 
$\sim10^{-5}$ by the COBE normalization \cite{normalization}, naively we might expect to find a 
requirement of $\Gamma^{-1} H^4 \ll 1$ instead.  This is equivalent to having many vacuum transitions 
per e-folding of inflation.  Indeed, we will ultimately show that about $10^4$ transitions per e-folding 
are needed in order to obtain sufficiently small density  perturbations.

In section 2 we describe our specific assumptions in more detail, as well as outline the method by which 
the density perturbations will be calculated.  This includes a demonstration that the amplitude of 
perturbations will be given in terms of the RMS fluctuations of the vacuum number at a point.  In section 
3 we give a description of the algorithm used in our simulations.  In section 4 we give our results, and 
we summarize and consider future directions in section 5.  Further details concerning the simulations and 
data analysis can be found in the appendices.


\section{Density Perturbations} \label{perturbations}

The generation of density perturbations in chain inflation is not entirely unlike the situation in 
ordinary slow-roll inflation.  In both cases, the perturbation modes can be understood as arising from 
$\bfx$-dependent field fluctuations, generated by the quantum mechanical randomness inherent in the 
microscopic evolution of each system.  Unless stated otherwise, we shall work in synchronous gauge, 
in which ``$t$'' labels proper time, and ``$\bfx$'' labels worldlines of free-falling observers, with 
the physical distance between two points being given at zeroth order in the perturbations by 
$a|\bfx-\bfy|$, where $a$ is the scale factor.  ``$t=0$'' may be taken to denote a spatially flat 
hypersurface in the distant past at a time at which our present horizon scale was located within a 
single bubble.   Note that we shall ignore the back-reaction of the metric perturbations on the 
evolution of the inflaton field, as this would be a higher order effect.

In the slowly rolling case, a given mode starts out far in the past in a coherently-evolving quantum 
ensemble -- a component of the scalar field vacuum in de Sitter space.  As the mode inflates outside 
of the horizon, the ensemble decoheres, leaving distinct ``classical'' field fluctuations \cite{Hawking} 
\cite{Starobinsky} \cite{GuthPi}.  The field then rolls down its potential deterministically throughout 
space, but its fluctuations cause nonzero relative timings 
\begin{equation}
\delta t(\bfx) = -\frac{\delta\phi(\bfx,t)}{\dot{\phi}_0(t)},
\end{equation}
where $\phi_0(t)$ is the spatially averaged field.  These in turn lead to metric and matter density 
fluctuations.  As observers looking at the cosmos today, we can then understand the fluctuations in 
the microwave sky and matter distribution as ultimately arising from the quantum-mechanical uncertainty 
of field values in the de Sitter vacuum.

Similar considerations also apply in chain inflation, although the dynamics are significantly different.  
As the tunneling field proceeds down the chain and the universe undergoes successive periods of 
percolation, the inherently random placement of bubble nucleation sites across spacetime causes certain 
patches of space to advance further along the chain than others, translating into timing fluctuations.  
As in the slow-roll case, these timing fluctuations are produced on scales smaller than the Hubble 
radius, and  evolution on trans-Hubble scales is again essentially deterministic.  The density 
perturbations thus produced can again be seen to be a product of quantum randomness, but now originating 
in the tunneling process instead of the decoherence of de Sitter vacuum modes.

Unfortunately, finding the statistical properties of the evolution of a given bubble network can be highly 
nontrivial.  The evolution is tied to the details of bubble wall collisions and the interaction with radiation 
from previous percolations.  If the tunneling rate varies wildly from vacuum to vacuum (as might be expected 
due to the exponential sensitivity of the tunneling rate to the form of the potential) then things can 
get very messy.  On the other hand it is possible to make a well defined set of simplifying assumptions 
so that a calculation of density perturbations becomes tractable.  We will assume the following four 
conditions:

\begin{enumerate}

\item  The size of the bubble immediately after tunneling is small compared to its characteristic 
       tunneling time, $\tau \equiv \Gamma^{-1/4}$.

\item  Upon collision, bubble walls dissipate quickly into radiation, on time scales shorter than 
       (or at least not much longer than) $\tau$.

\item  Tunneling which skips steps in the chain is sufficiently suppressed that it can be ignored.

\item  All steps in the chain have identical tunneling events.  This includes tunneling rates (perhaps 
       effective thermal tunneling rates), as well as the total vacuum energy lost at each step.

\end{enumerate}

As noted in the introduction, this final assumption is the only obvious way to ensure that the final 
spectrum is scale invariant, and it also greatly simplifies our calculation method.  To the extent 
that our assumptions are not realistic for a given scenario, our results may be viewed as a baseline 
for more complicated treatments and/or construction of more advanced simulation methods.

Note that as the universe expands, radiation left over from earlier percolations is inflated away.  
However, it takes about an e-folding of inflation for a given quantity of radiation to be noticeably 
diluted.  This implies that there is always present a bath of radiation, which has an average energy 
density given by the total vacuum energy lost in about one Hubble time.  Eventually the vacuum energy 
drops below the energy density in this radiation, and inflation comes to an end.  Note that assumptions 
1 through 4 need only be valid up to an epoch about 30 to 60 e-foldings before the end of inflation 
in order to satisfy the observational constraints from the CMB (i.e., up to the epoch when our current Hubble 
scale exited the horizon). If they are always true, on the other hand, then in fact we can make an estimate 
for the reheating temperature at the end of inflation:  In this case the Hubble parameter may be written as 
\begin{equation}
H(t) \sim \frac{\sqrt{\rho_r(t) + \rho_V(t)}}{M_{pl}},
\label{eq:H}
\end{equation}
where $\rho_r$ is the average radiation energy density, and $\rho_V$ is the vacuum energy.  Since we 
have about one e-folding worth of radiation at any given time, $\rho_r$ satisfies
\begin{equation}
\rho_r(t) \sim \frac{\epsilon}{H(t)\tau},
\label{eq:rho}
\end{equation}
with $\epsilon$ representing the vacuum energy lost in each step of the chain.\footnote{The naive 
expectation that the vacuum energy goes like $\rho_V \sim \rho_{V,i} - \epsilon t / \tau$ will be 
confirmed later in our simulations.}  This implies the reheating temperature will be about  

\begin{equation}
T_R \sim \left( \frac{M_{pl} \epsilon}{\tau}\right)^{1/6}.
\label{eq:T}
\end{equation}

Finally, we note that in what follows, we will treat the Hubble parameter as roughly constant during 
the period of inflation responsible for generating the modes of the density perturbations probed by 
observations of the CMB.  This should be a good approximation, since this period is expected to be 
fairly short compared to the epoch of inflation  remaining after our present Hubble volume exited 
the horizon.  This implies that during the era of production of the observed density perturbations, 
the vacuum energy could only have changed by a small fraction.

With our setup and assumptions now in hand, we may next lay out a systematic program for calculating 
the size of the density perturbations in this model.  In order to preserve the analogy with slow roll 
inflation, we will denote the vacuum number as a function of position by $\phi({\bf x},t)$.  Note, 
however, that with this definition $\phi$ is a dimensionless field.

We begin by constructing the spatial two point correlator of $\phi$ fluctuations, defined as the 
expectation value $\langle(\phi(\bfx,t) - \vev{\phi(t)})\,(\phi(\bfy,t) - \vev{\phi(t)})\rangle = 
\langle\phi(\bfx,t)\,\phi(\bfy,t)\rangle - \vev{\phi(t)}^2 $, where we average over an ensemble of 
universes (or physically remote regions in a single universe).  Note that the zero-mode of this 
quantity is unobservable, as it will incorporate fluctuations of $\phi$ inaccessible to any single 
observer viewing only a particular local patch.  Other Fourier modes of this object do have a 
well-defined physical meaning, however;  they represent the power in each field fluctuation mode, 
and will ultimately translate into the power of density fluctuations observed in the CMB.  

We will now demonstrate that the two point correlator of $\phi$ fluctuations may be expressed on large 
scales in terms of the standard deviation of $\phi$ at a {\it single} point, plus a constant term 
which contributes only to the zero-mode.  In order to see this, let us consider the evolution of the 
correlator along two worldlines at positions ${\bf x}_A$ and ${\bf x}_B$ which are separated by a 
physical trans-Hubble distance $L$ at the time $t$.  There will be an earlier time 
\begin{equation}
t_\tau(L) \sim t - H^{-1} \log \frac{L}{\tau}
\label{eq:tautime}
\end{equation}
before which the two worldlines were closer together than the characteristic size of the percolating 
bubbles, and therefore typically inside the same bubble.  Let us call the common field value at this 
time $\phi(t_\tau(L))$.  As the worldlines are driven apart by the Hubble flow, they depart from their 
mutual bubble and experience distinct but correlated evolutions until a time $t_H$ shortly after they 
leave each other's horizons.  Their evolution subsequently decouples.  The complete evolution therefore 
breaks up into three distinct periods:

\begin{enumerate}

\item $t < t_\tau$:  Completely correlated evolution.  At the end of this period, worldlines $A$ and 
      $B$ share some common value $\phi(t_{\tau}(L))$.

\item $t_\tau < t < t_H$:  Partially correlated evolution.  Each worldline acquires a correlated advance 
      in field value $\delta_A$ and $\delta_B$, respectively, with $\vev{ \delta_A \, \delta_B} - 
      \vev{\delta}^2 > 0 $.

\item $ t > t_H$:  Completely uncorrelated evolution.  Each worldline acquires an uncorrelated advance 
      in field value $\Delta_A$ and $\Delta_B$, respectively, with $\vev{\Delta_A \, \Delta_B} - 
      \vev{\Delta}^2 \simeq 0$ and $\langle \Delta_{A/B} \, \delta_{A/B} \rangle \simeq 
      \langle \Delta \rangle \; \langle \delta \rangle $.

\end{enumerate}

\begin{figure}[ht]
\begin{center}
\epsfxsize=0.44\textwidth\epsfbox{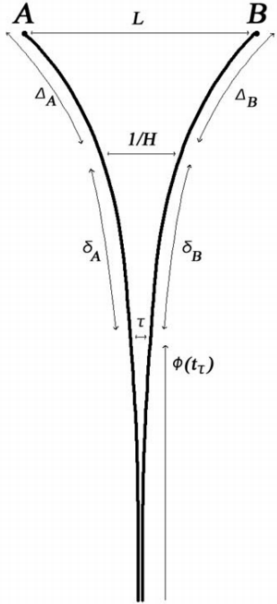}
\caption{Schematic representation of the evolution of the 2-point correlator of $\phi$ fluctuations. 
         $\phi(t_{\tau})$ represents the totally correlated evolution at scales less than $\tau$, 
         $\delta$'s represent the partially correlated evolution on scales between $\tau$ and $1/H$ and 
         $\Delta$'s represent the totally uncorrelated evolution on scales greater than $1/H$.}
\label{fig:correlatorEvolution}
\end{center}
\end{figure}

The situation is illustrated in figure \ref{fig:correlatorEvolution}.  Representing $\phi_A(t) = 
\phi(t_{\tau}(L)) + \delta_A + \Delta_A$, and similarly for $B$, we obtain

\begin{equation}
 \langle\phi_A(t)\,\phi_B(t)\rangle - \vev{\phi(t)}^2
= \left[ \langle \phi(t_{\tau}(L))^2 \rangle - \langle \phi(t_{\tau}(L)) \rangle^2 \right] + 
  \left[ \langle \delta_A \, \delta_B \rangle - \langle \delta \rangle^2 \right] .
\end{equation}

If we consider a different value of $L$ at the same $t$, the argument repeats.  However, we are assuming 
that the system has nearly-constant $\tau$ and nearly-constant $H$, and that its evolution is essentially 
insensitive to absolute time.  In particular, period \#2 above -- evolution from $\tau$ separation to 
Hubble separation -- is independent of $L$, and so should look essentially {\it identical}  for all 
pairs of points that end up outside the horizon by the time $t$.  The generic trans-horizon spatial 
correlator then takes the form
\begin{equation}
  \langle\phi_A(t)\,\phi_B(t)\rangle - \vev{\phi(t)}^2
 =  \left[\langle \phi(t_{\tau}(L))^2 \rangle - \langle \phi(t_{\tau}(L)) \rangle^2\right]  + const,
\label{eq:generalcorrelate}
\end{equation}
with the constant referring to terms independent of $L$.  As claimed, the problem of determining the 
two-point correlator of $\phi$ thus translates into the problem of determining its RMS fluctuations at 
a point as a function of time.  

If the system is truly scale-invariant, then different slices of time evolution should behave 
statistically in the same way, accruing identical fluctuations.\footnote{There will be very early 
times before which the scale of our entire observable universe was contained within a distance scale 
less than $\tau$, and during which the Hubble parameter could have been appreciably higher than its 
later size.  Such periods will contribute to $\phi_{RMS}(t)^2$ in ways not proportional to $t$, but 
their effect will be $L$-independent on scales within our present horizon.}  This implies that 
$\vev{\phi(t)}$ and the RMS fluctuations of $\phi$ take the form
\begin{equation}
\vev{\phi(t)} =  \alpha(H\tau) \; H t,
\label{eq:alphadef}
\end{equation}
\begin{equation}
\phi_{RMS}(t)^2  \equiv \langle \phi(t)^2 \rangle - \langle \phi(t) \rangle^2 = \beta(H\tau) \; H t,
\label{eq:betadef}
\end{equation}
with $\alpha$ and $\beta$  yet-to-be-determined functions of the nucleation density. Extra factors of 
$H$ have been inserted for convenience of normalization.  It now follows from equations \ref{eq:tautime}, 
\ref{eq:generalcorrelate} and \ref{eq:betadef} that
\begin{equation}
\langle \phi(\bfx,t) \, \phi(\bfy,t) \rangle - \vev{\phi(t)}^2 = -\beta(H\tau) \; \log|\bfx-\bfy| + const.
\label{eq:logform}
\end{equation}
We repeatedly verify the behavior \ref{eq:alphadef} and  \ref{eq:betadef}  in our numerical simulations, 
and explicitly demonstrate the logarithmic scaling of the correlator for several values of $\tau$ and 
$t$.  We describe the details of these simulations in the subsequent sections.  Now, taking the Fourier 
transform of the correlator (and invoking spatial homogeneity) we obtain 
\begin{equation}
\langle \delta\phi_\bfk(t) \, \delta\phi_{\bfk^\prime}(t) \rangle = 
(2\pi)^3 \delta^3(\bfk-\bfk^\prime) \frac{2\pi^2}{k^3}\;\beta(H\tau) ,
\end{equation}
which is the desired scale-invariant form (here the $\bfk$'s label comoving wavenumbers). This yields 
a power spectrum for $\delta \phi$ given by \footnote{Recall that the power spectrum for a field $g$ is 
defined by $\vev{g_\bfk \, g_{\bfk^\prime}}  =  (2\pi)^3 \delta^3(\bfk-\bfk^\prime) \frac{2\pi^2}{k^3} 
\; {\cal P}_g $.}

\begin{equation}
{\cal P}_{\delta \phi} = \beta(H\tau).
\label{eq:spectrum}
\end{equation}

Now, when a mode of the density perturbations $\frac{\delta \rho_\bfk}{\rho}$ re-enters the horizon 
during matter domination, its size on a {\it comoving} hypersurface (one on which the momentum density 
vanishes) is related to the spatial curvature $R_\bfk^{(3)}$ on that surface, through the relation 
\cite{LR}
\begin{equation}
\frac{\delta\rho_\bfk}{\rho} = \frac25 {\cal R}_\bfk.
\end{equation}
Here ${\cal R}_\bfk$ is defined by 
\begin{equation}
{\cal R}_\bfk \equiv \frac{1}{4} \left(\frac{a}{k}\right)^2 R^{(3)}_\bfk.
\end{equation}
The observations of the CMB by COBE then constrain the current power spectrum  ${\cal P}_{\cal R}$, 
for modes of order the Hubble radius today.  The value  obtained is \cite{normalization}
\begin{equation}
\left( {\cal P}_{\cal R}^{COBE} \right)^{1/2} = 4.8\times10^{-5} .
\label{eq:COBE}
\end{equation}
This quantity may be related to the synchronous gauge ${\cal P}_{\delta \phi}$ which appears in equation 
\ref{eq:spectrum}, evaluated at horizon {\it exit}.   To see this, consider the epoch when a perturbation 
mode first exits the horizon during inflation.   In synchronous gauge, the fluctuations in the vacuum 
number yield fluctuations in the spatial curvature through Einstein's equations of order ${\cal R}_0 
\sim \frac{\delta \rho_V}{\rho_V} \sim -\epsilon\frac{\delta \phi}{\rho_V}$, with the vacuum energy 
$\rho_V$ taking the form $\rho_V= \rho_{V,i} - \epsilon \phi$.  We then must make a transformation to 
comoving coordinates and see how the spatial curvature changes. 

In slow roll inflation, comoving hypersurfaces correspond on trans-horizon scales to surfaces of constant 
field value \cite{LR}, and the same will be true in chain inflation.  The momentum density of a 
chain-inflating field is localized inside bubble walls and  is constantly downconverting into radiation.  
It is therefore the radiation which will dominate the momentum density of the system, and we must choose 
a frame in which that momentum averages out to zero.  This will happen in a spatial slicing where 
tunneling to a particular vacuum is occurring roughly instantaneously across space, and in which bubbles 
of this vacuum collide with roughly equal sizes on average.  In any different slicing, these vacuum 
transitions occur at different times across space, and bubbles of a given vacuum will collide with 
systematically different sizes, imparting net average momentum to the daughter radiation.  In order 
that the successive generations of radiation not encode this momentum, it is then clear that comoving 
hypersurfaces, on large scales, will be slices of constant $\phi$.  The coordinate transformation to 
comoving coordinates, with a time shift of $\delta t(\bfx)$, then induces an additional contribution 
to the curvature perturbation of size $\Delta {\cal R}_\bfk = H\;\delta t_\bfk = 
-H \frac{\delta \phi_\bfk}{\dot{\phi}}$.  Since the fractional change in $V$ during one Hubble time 
is small, this additional term added to ${\cal R}$ in making the transformation is actually much 
larger than the original contribution ${\cal R}_0$ that we started with.  Thus when a mode of the 
curvature perturbation exits the horizon during inflation, it has a size in comoving coordinates 
given by \cite{LR}
\begin{equation}
{\cal R}_\bfk =  -H \frac{\delta \phi_\bfk}{\dot{\phi}}.
\label{eq:Rf}
\end{equation}
Furthermore, it is known that the size of ${\cal R}_\bfk$ in comoving gauge will be time independent 
outside of the horizon \cite{LR}.  We may thus use expression \ref{eq:Rf} to compare our result 
\ref{eq:spectrum} for the $\phi$ power spectrum with the COBE normalization \ref{eq:COBE}.  We obtain
\begin{equation}
\left({\cal P}_{\cal R}^{COBE}\right)^{1/2} = \frac{\sqrt{\beta(H\tau)}}{\alpha(H\tau)} ,
\label{eq:result}
\end{equation}
where $H$ and $\tau$ are to be evaluated at horizon exit.  The following section is thus devoted to 
determining the functions $\alpha(H\tau)$ and $\beta(H\tau)$ characterizing the statistical evolution 
of the field at a fixed point in space.


\section{Fluctuations along one Worldline} \label{fluctuations}

Consider a point $(\bfx,t)$ in de Sitter space.  We will be interested in this point's entire past light 
cone back to an initial time ``$t=0$'' at which we will say the field value was at step ``0''.   We would 
like to calculate the statistical properties of the vacuum number occupied by $(\bfx,t)$.  Note that our 
assumption of a uniform initial condition was an arbitrary choice:  The random fluctuations in the 
tunneling process will wash out the effects of the initial condition in around a Hubble time.  What we 
will be interested in is the {\it asymptotic} behavior of $\vev{\phi(\bfx,t)} = \vev{\phi(t)}$  and 
$\phi_{RMS}(\bfx,t)^2 = \phi_{RMS}(t)^2$ for times much larger than $1/H$.

Since we have assumed that the tunneling rate is independent of the step number in the chain, the 
probability per unit volume per unit time for a bubble nucleation event is a fixed number, $\Gamma$.  
The probability of there being $N$ bubble nucleations in the past lightcone of $(\bfx,t)$ is then 
given by the Poisson distribution
\begin{equation}
Pr(N,V) = \frac{1}{N!}(\Gamma V)^N e^{-\Gamma V},
\label{eq:poisson}
\end{equation}
where $V$ is the 4-volume in the past lightcone back to the initial surface.  The mean number of bubble 
nucleations is $\Gamma V$.

\begin{figure}[t]
\begin{center}
\epsfxsize=0.4\textwidth\epsfbox{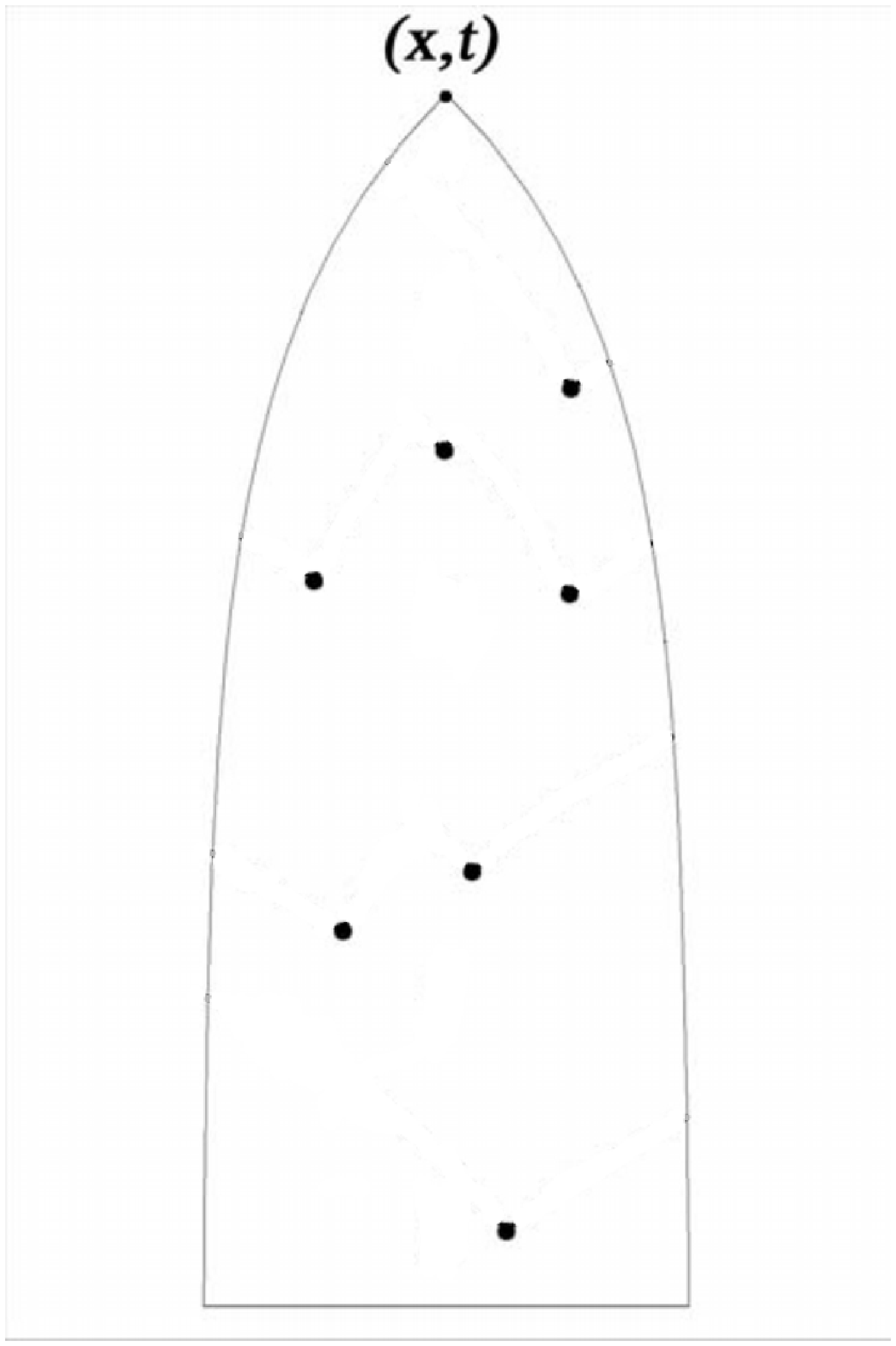}
\caption{The past light cone of the point ({\bf x}, t) in de Sitter space, containing a randomly 
         generated set of 7 bubble nucleations. The horizontal axis represents physical distances.}
\label{fig:dots}
\end{center}
\end{figure}

A sample history for $(\bfx,t)$ with $N$ bubbles may then be represented by distributing the bubbles at 
random throughout the past lightcone, as shown in figure \ref{fig:dots}.  The question then is, given a 
configuration of nucleation sites in this way, what is the resulting value of the field at $(\bfx,t)$?

Each nucleation site will have the boundary of its future lightcone traced out by its own expanding 
bubble wall.  As shown in figure \ref{fig:dotsAndWalls}, we could in principle determine the vacuum 
number at $(\bfx,t)$ by following along the sequence of bubble nucleations and wall collisions.  This 
picture is rather complicated, and inefficient for the purpose of computer simulations.  This is 
essentially because calculating the final field value in this way requires keeping track of  the vacuum 
numbers for a continuum of points, and then evolving them with time.  In order to avoid doing this, we 
will instead 
use the following simpler picture:

\begin{figure}[t]
\begin{center}
\epsfxsize=0.4\textwidth\epsfbox{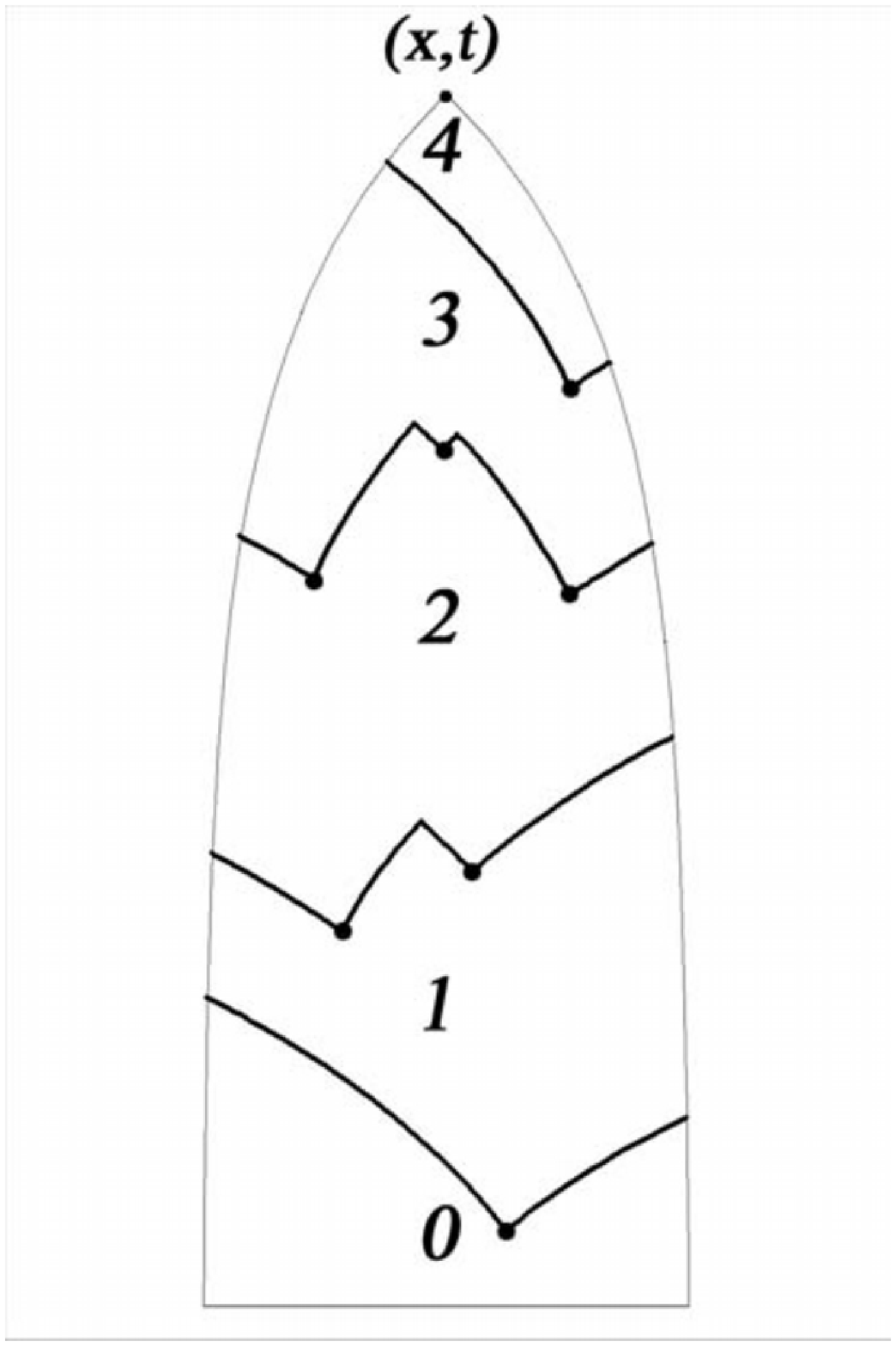}
\caption{The bubble nucleations from figure \ref{fig:dots}, shown again with their associated bubble 
         walls and vacuum numbers.}
\label{fig:dotsAndWalls}
\end{center}
\end{figure}

Label each nucleation site ``$i$'' with an integer $\phi(i)$, corresponding to the vacuum number in the 
interior of the newly formed bubble.  Also, let the past lightcone of a spacetime point or a nucleation 
site be denoted by $L(\bfx,t)$ or $L(i)$, respectively.

Then the vacuum number at $(\bfx,t)$ is given by 
\begin{equation}
\phi(\bfx,t) = \displaystyle\max_{j \in L(\bfx,t)} \phi(j),
\label{eq:vac}
\end{equation}
where we take the maximum of vacuum numbers over all nucleation sites in the past lightcone of 
$(\bfx,t)$.  Here we have ignored the vanishing probability that there is a bubble nucleation at the 
point $(\bfx,t)$ itself.  The vacuum number at an actual nucleation site is given by
\begin{equation}
\phi(i) = \displaystyle\max_{j \in L(i)} \phi(j) +1 .
\label{eq:vac2}
\end{equation}
Expression \ref{eq:vac} gives us the vacuum number at the chosen point for a particular configuration 
of bubble nucleations.  The statistical properties of $\phi(\bfx,t)$ will then be determined by summing 
up over contributions from all such configurations.  Obtaining analytic expressions for the resulting 
mean and variance for $\phi(\bfx,t)$ is a very hard problem, and for this reason we will resort to 
computer simulations rather than  analytic analysis from this point forward.  In fact, the simple 
characterization of vacuum numbers given by \ref{eq:vac} and \ref{eq:vac2} immediately suggests a 
useful algorithm for this purpose, which we now describe.  In what follows, we will ignore certain 
approximations that were adopted in the code in order to speed up running time.  A complete discussion 
of these approximations will be presented in appendix \ref{appendixA}.  

To find the vacuum number at $(\bfx,t)$ for a specific history of nucleations, we do the following:
$\newline$
\begin{enumerate}

\item Time order the bubble nucleation sites, and put them in a list.

\item Associate a vacuum number with each site in the list, initialized to zero.  This number will 
      be replaced with its correct value as the simulation progresses.

\item Go through the nucleation sites in order from earliest to latest.  At each site, set the vacuum 
      number according to equation \ref{eq:vac2}, by searching over all prior sites.

\item After the vacuum numbers for all sites have been correctly set, use equation \ref{eq:vac} to 
      find the vacuum number at $(\bfx,t)$.

\end{enumerate}
$\newline$
Now, the goal of the simulations is ultimately to calculate $\vev{\phi(\bfx,t)}$ and 
$\phi_{RMS}(\bfx,t)^2$.  To find these expectation values, we have employed a monte carlo method:  
Dots are distributed at random throughout the past light cone of $(\bfx,t)$, in numbers given by the 
Poisson distribution $\ref{eq:poisson}$.  We then find the resulting $\phi(\bfx,t)$ and repeat the 
process many times, determining $\vev{\phi(\bfx,t)}$ and $\phi_{RMS}(\bfx,t)^2$ statistically.  

The results of our simulations are presented in the next section.


\section{Results}  \label{results}

\begin{figure}[p]
\begin{center}
\epsfxsize=0.7\textwidth\epsfbox{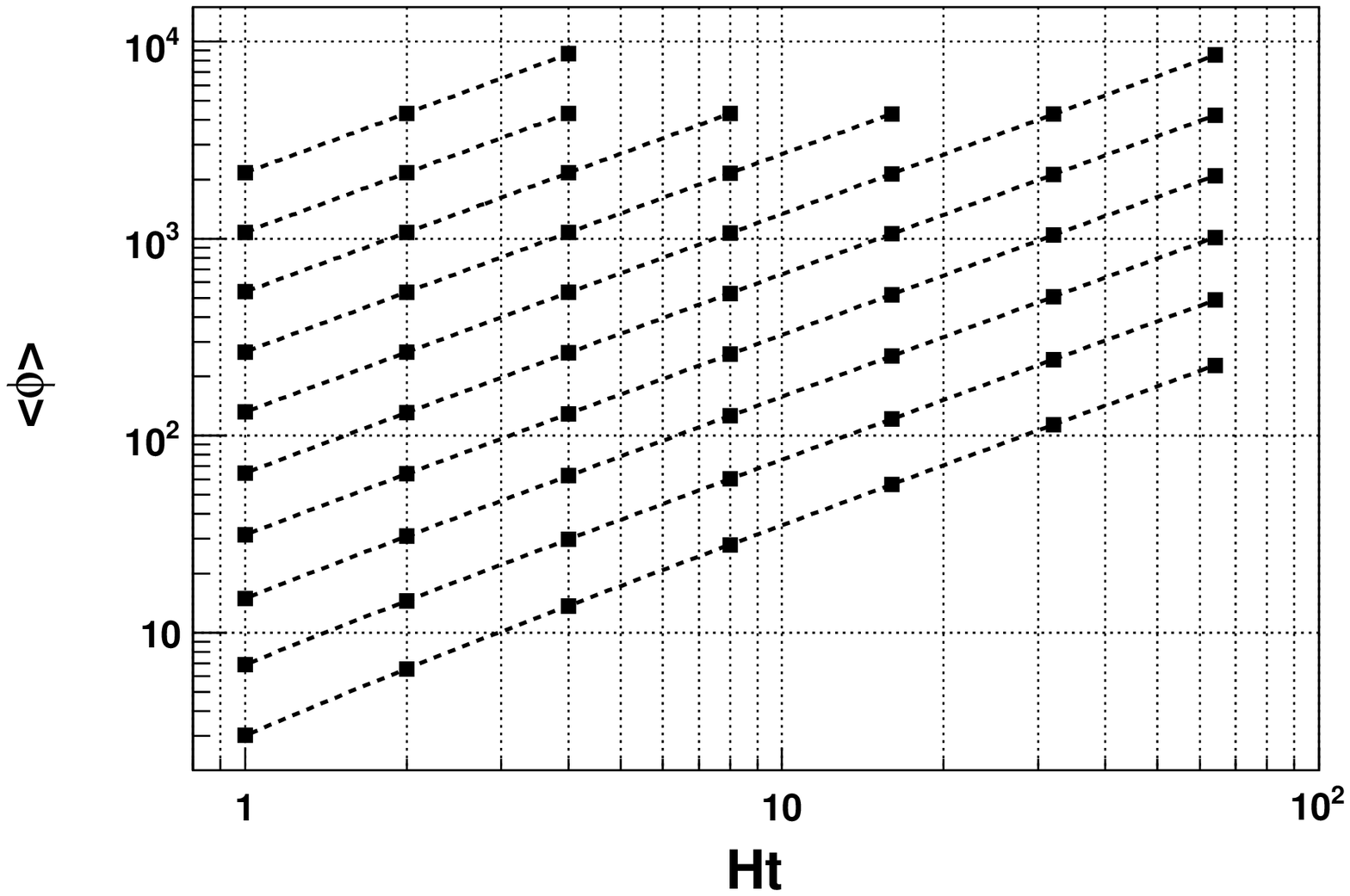}
\caption{$\vev{\phi}$ graphed as a function of $t$ in 1+1 dimensions with different $\tau$'s.  Points 
         of each $\tau$ are connected by best-fit straight lines in untransformed coordinates.  From 
         bottom to top, $(H\tau)^{-1} = 3,6,12,\ldots,1536$.}
\label{fig:meanVsT2D}
\vspace{0.5in}
\epsfxsize=0.7\textwidth\epsfbox{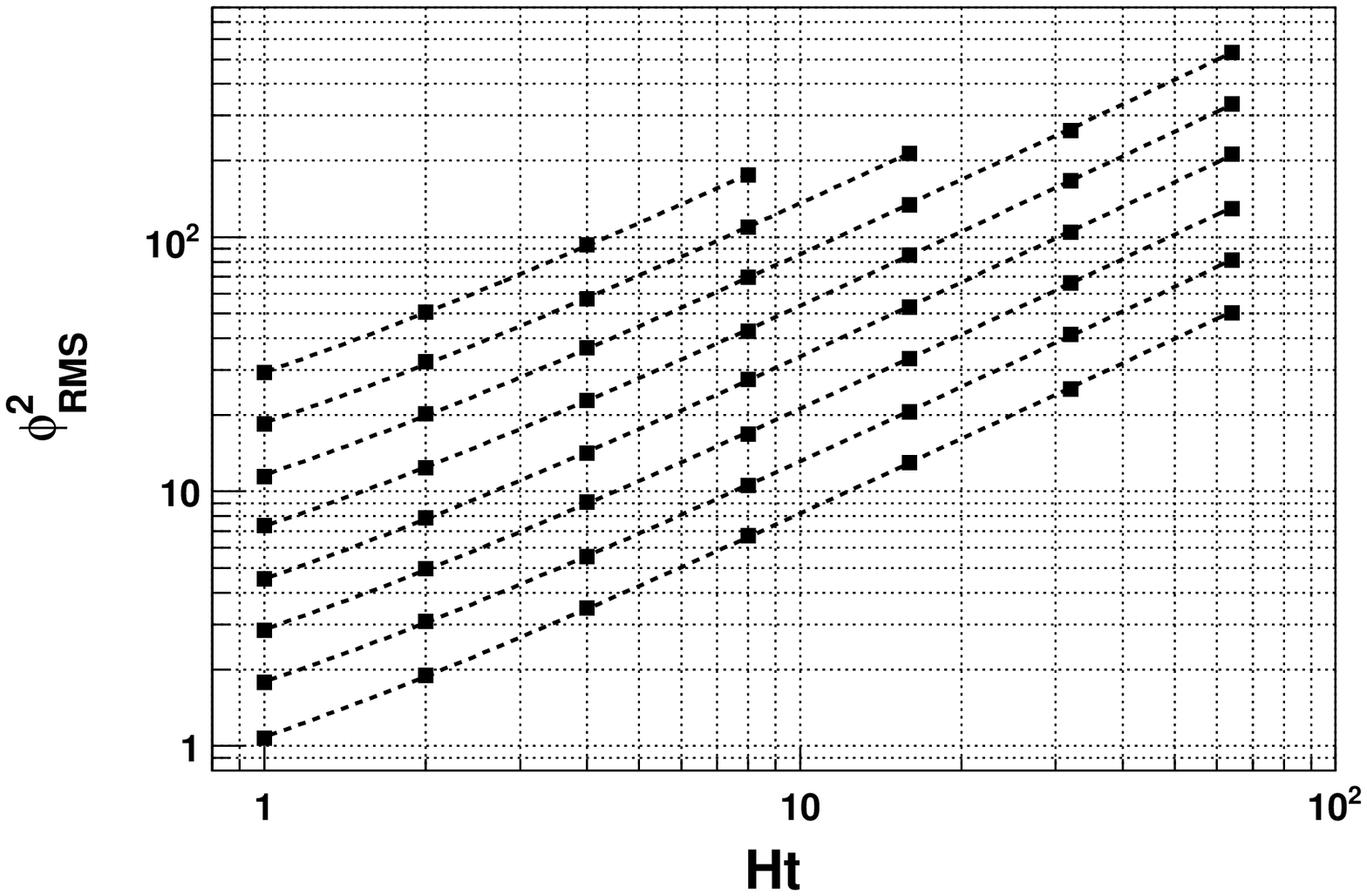}
\caption{$\phi_{RMS}^2$ graphed as a function of $t$ in 1+1 dimensions with different $\tau$'s.  Points 
         of each $\tau$ are connected by best-fit straight lines in untransformed coordinates.  From 
         bottom to top, $(H\tau)^{-1} = 3,6,12,\ldots,384$.}
\label{fig:sigmaVsT2D}
\end{center}
\end{figure}

\begin{figure}[p]
\begin{center}
\epsfxsize=0.7\textwidth\epsfbox{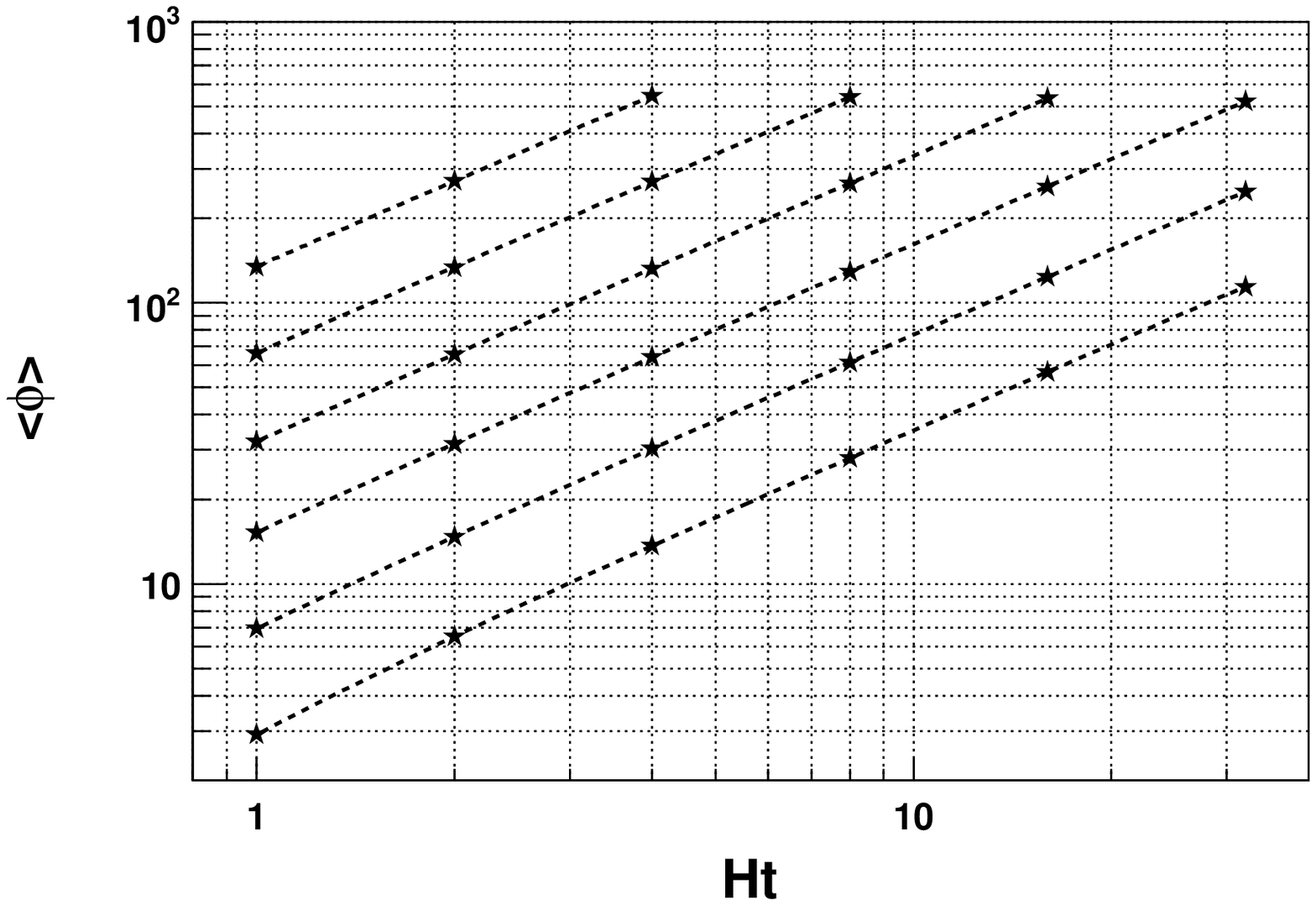}
\caption{$\vev{\phi}$ graphed as a function of $t$ in 2+1 dimensions with different $\tau$'s.  Points 
         of each $\tau$ are connected by best-fit straight lines in untransformed coordinates.  From 
         bottom to top, $(H\tau)^{-1} = 3,6,12,\ldots,96$.}
\label{fig:meanVsT3D}
\vspace{0.5in}
\epsfxsize=0.7\textwidth\epsfbox{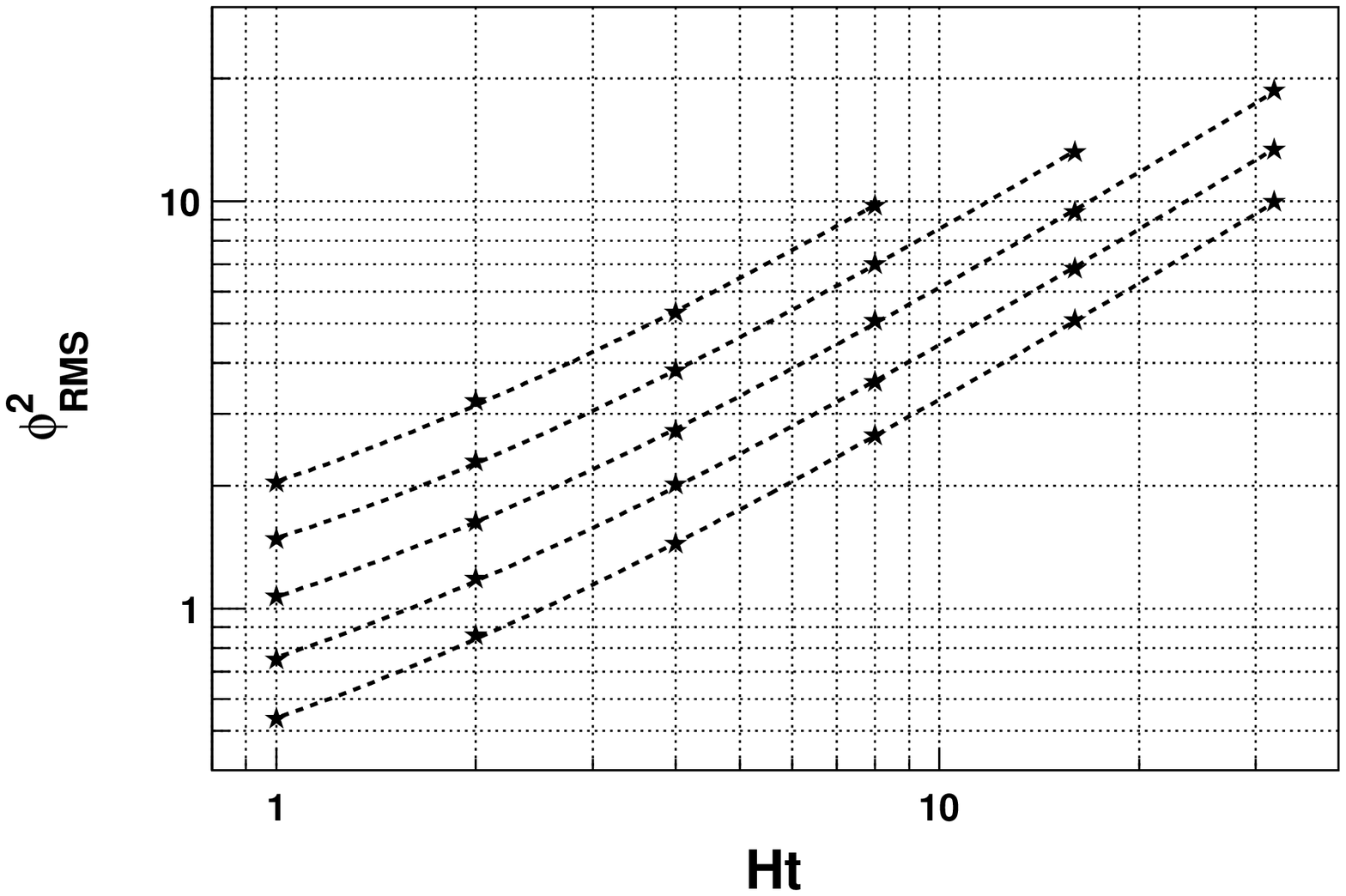}
\caption{$\phi_{RMS}^2$ graphed as a function of $t$ in 2+1 dimensions with different $\tau$'s.  Points 
         of each $\tau$ are connected by best-fit straight lines in untransformed coordinates.  From 
         bottom to top, $(H\tau)^{-1} = 3,6,12,\ldots,48$.}
\label{fig:sigmaVsT3D}
\end{center}
\end{figure}

\begin{figure}[p]
\begin{center}
\epsfxsize=0.7\textwidth\epsfbox{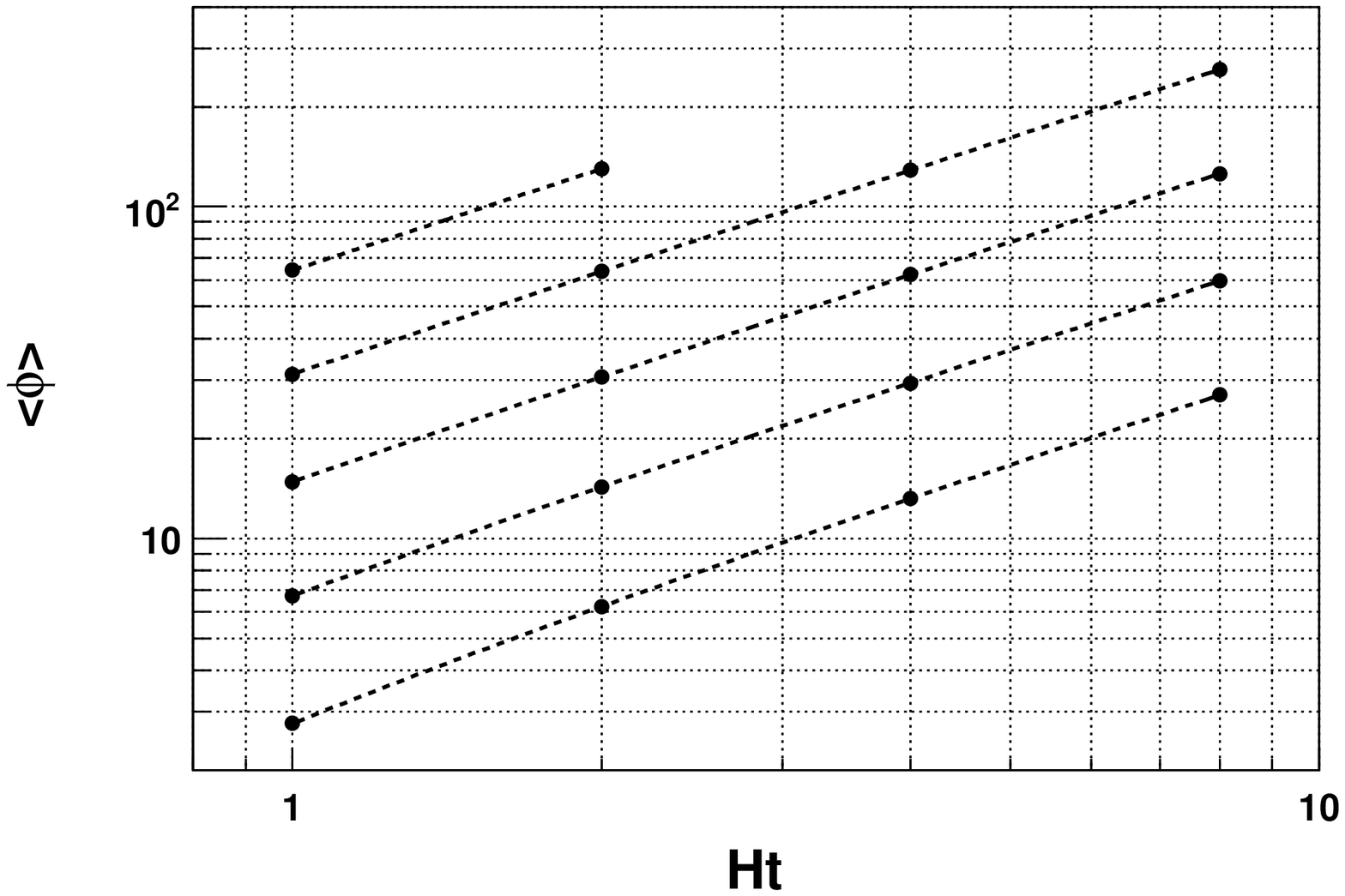}
\caption{$\vev{\phi}$ graphed as a function of $t$ in 3+1 dimensions with different $\tau$'s.  Points 
         of each $\tau$ are connected by best-fit straight lines in untransformed coordinates.  From 
         bottom to top, $(H\tau)^{-1} = 3,6,12,\ldots,48$.}
\label{fig:meanVsT4D}
\vspace{0.5in}
\epsfxsize=0.7\textwidth\epsfbox{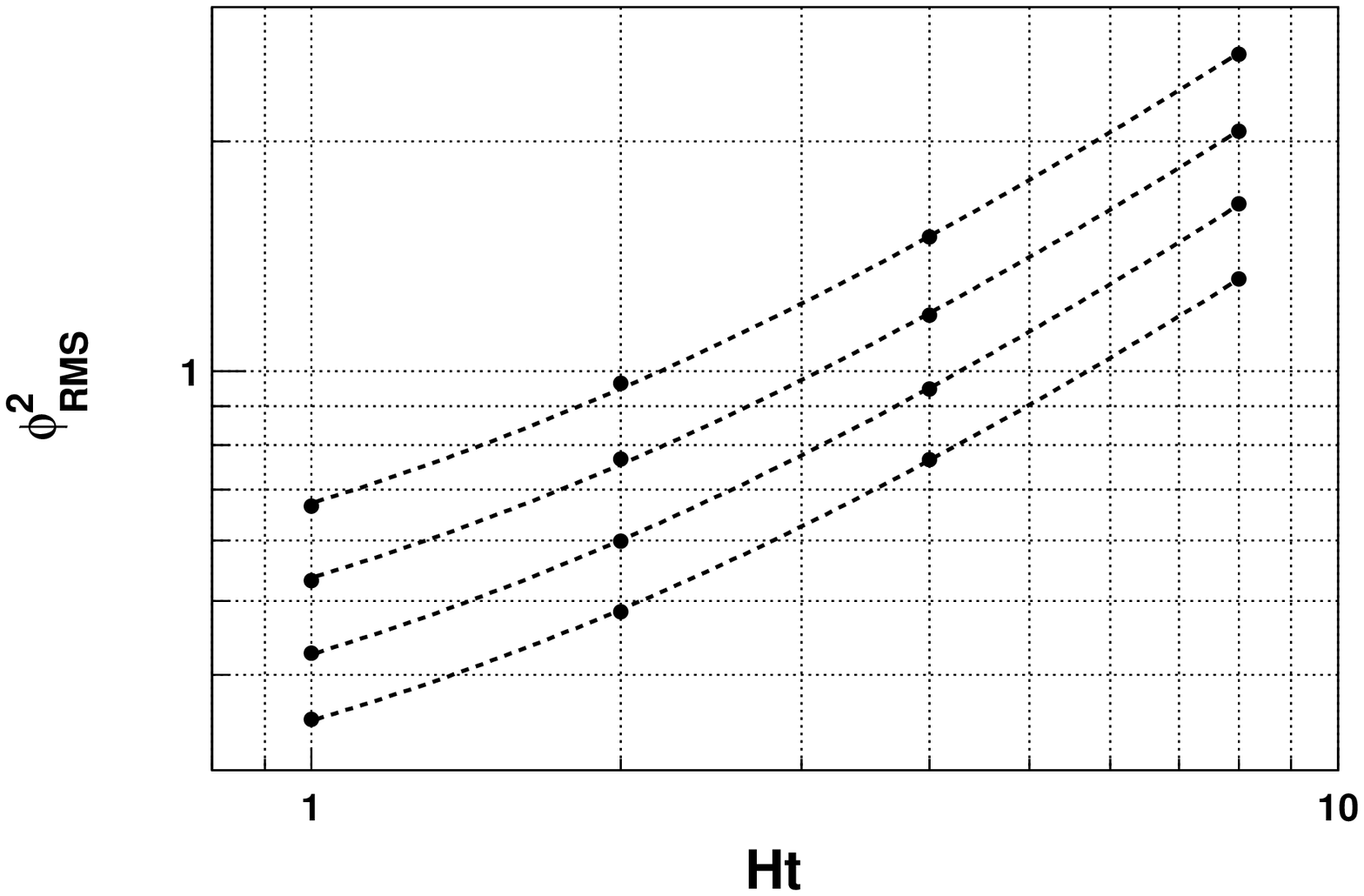}
\caption{$\phi_{RMS}^2$ graphed as a function of $t$ in 3+1 dimensions with different $\tau$'s.  Points 
         of each $\tau$ are connected by best-fit straight lines in untransformed coordinates.  From 
         bottom to top, $(H\tau)^{-1} = 3,6,12,24$.}
\label{fig:sigmaVsT4D}
\end{center}
\end{figure}

\begin{figure}[p]
\begin{center}
\epsfxsize=0.7\textwidth\epsfbox{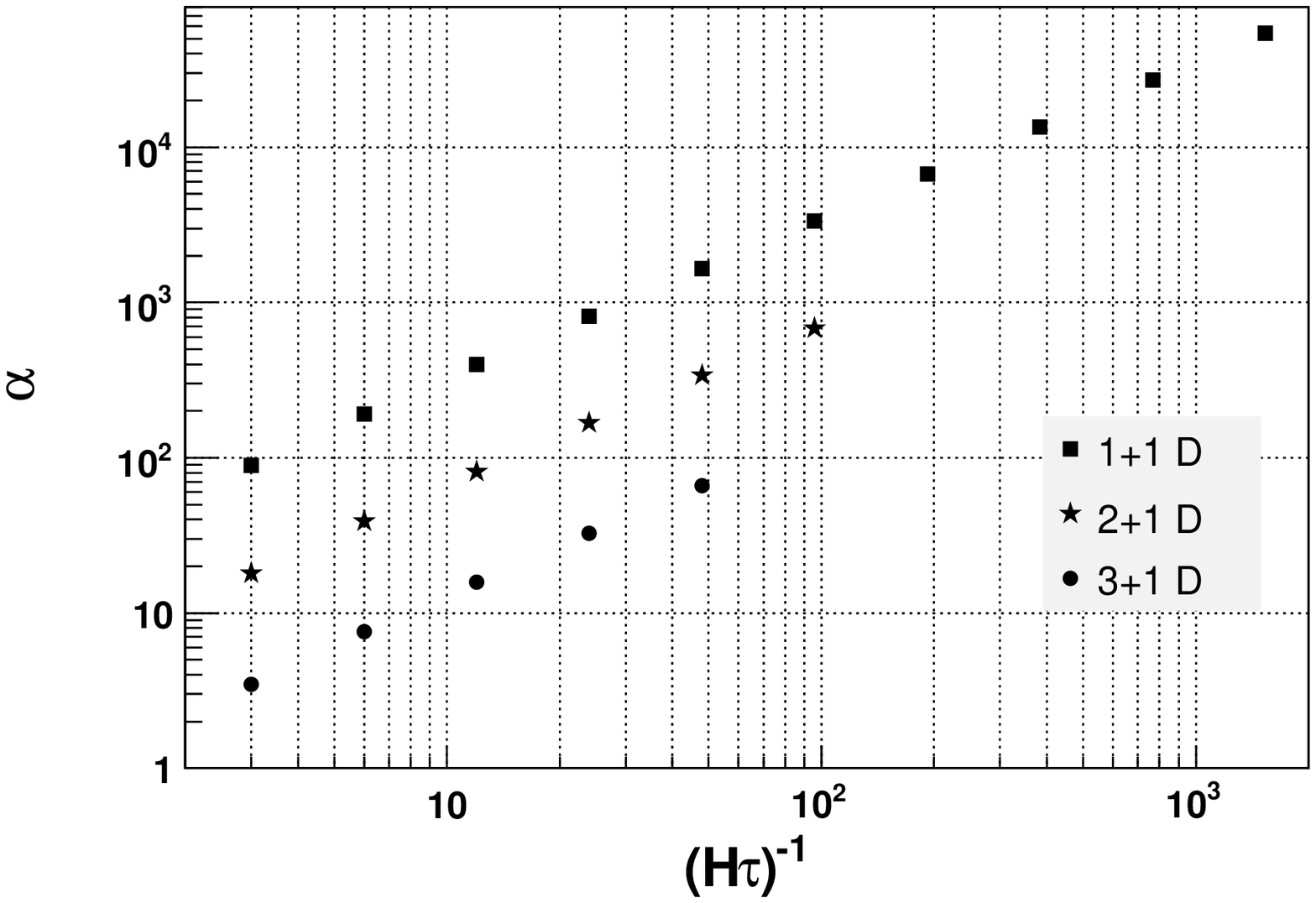}
\caption{The function $\alpha$, as described in the text, plotted versus $(H\tau)^{-1}$ for all 3 
         dimensions.  For clarity, the points for 1+1 dimensions and 2+1 dimensions have been 
         offset vertically by factors of 25 and 5, respectively.}
\label{fig:alpha}
\vspace{0.5in}
\epsfxsize=0.7\textwidth\epsfbox{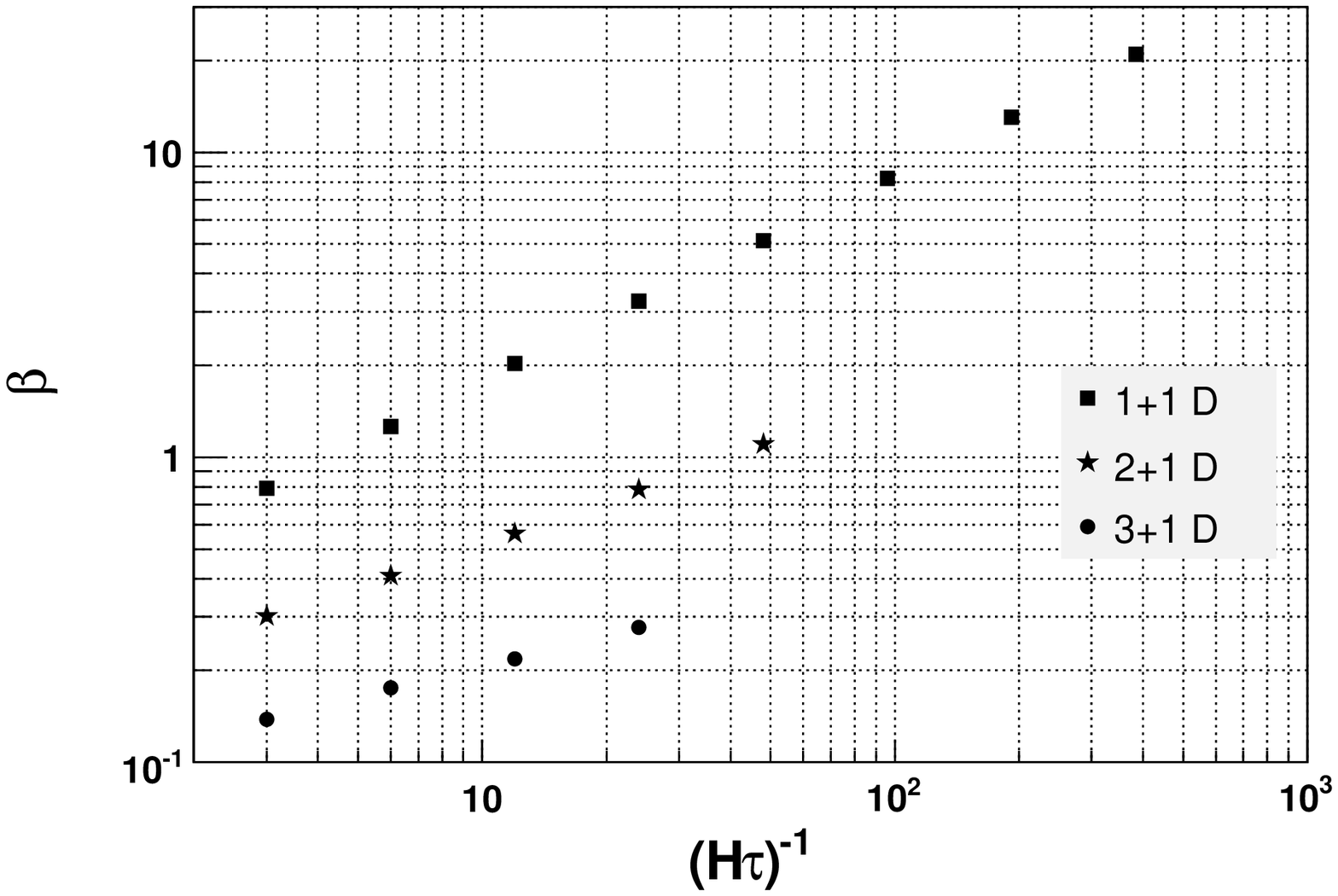}
\caption{The function $\beta$, as described in the text, plotted versus $(H\tau)^{-1}$ for all 3 
         dimensions.}
\label{fig:beta}
\end{center}
\end{figure}

We ran the simulations in 1+1, 2+1 and 3+1 dimensions (analytic expressions are trivially obtained in 
0+1 dimensions), typically for 25k events for each configuration.\footnote{Fractional statistical errors 
for $\vev{\phi}$ are much smaller than those for, $\phi_{RMS}^2$.  We therefore also ran smaller numbers 
of events with more calculationally-intensive parameters, which gave good resolution on the former but 
not the latter.  For those points, the results for $\phi_{RMS}^2$ are omitted from the analysis.}  We 
universally find the expected result that both $\vev{\phi(t)}$ and $\phi_{RMS}(t)^2$ increase linearly 
with time.  Figures \ref{fig:meanVsT2D} through \ref{fig:sigmaVsT4D} show plots of our data for each 
dimensionality, in log-log coordinates.  Each set of measurements of $\vev{\phi(t)}$ and 
$\phi_{RMS}(t)^2$ for the individual $\tau$ are fit with straight lines (which can appear curved in 
the log-log coordinates of the plots due to nonzero intercept).  The error bars are typically 
$(6.3\times10^{-3})\phi_{RMS}$ for $\vev{\phi}$, and $(8.9\times10^{-3})\phi_{RMS}^2$ for $\phi_{RMS}^2$, 
which are both much smaller than the data points on the plots.  Figures \ref{fig:alpha} and 
\ref{fig:beta} show plots of the slopes $\alpha$ and $\beta$ derived from these fits.

Using equation \ref{eq:result}, we immediately find that nowhere in the range of $\tau$'s explored 
by our simulation does chain inflation produce an acceptable power of density perturbations, placing a 
modest lower-limit $(H\tau)^{-1} \gg 24$.  In order to estimate the value which reproduces the COBE 
normalization, we perform an extrapolation of our data based on the assumption that, for asymptotically 
large values of $(H\tau)^{-1}$,
\begin{equation}
\alpha(H\tau) \sim A(H\tau)^m
\label{eq:vacform}
\end{equation}
\begin{equation}
\beta(H\tau) \sim B(H\tau)^n.
\label{eq:sigmaform}
\end{equation}
We find that this ansatz is indeed consistent with the data in all dimensions, and in particular in lower 
dimensions where we ran to larger values of $(H\tau)^{-1}$.  Details regarding the extrapolation method 
can be found in appendix \ref{appendixA}.  The results for the parameters characterizing the asymptotic 
behavior are summarized in table \ref{tab:results}.  Using the above expressions for $\alpha$ and $\beta$ 
in combination with equation \ref{eq:result}, we then obtain the estimate
\begin{equation}
(H\tau)^{-1} = (2.2\pm0.8)\times10^4.
\end{equation}
This number, when multiplied by the factor $A \sim 1.5$ gives the required number of vacuum transitions 
per e-folding of inflation.

\begin{table}[t]
\begin{center}
\begin{tabular}{|c|c|c|c|c|}  \hline 
                   &         A         &         m         &         B         &        n    
\\ \hline
 0+1               &   $1$              &   $-1$             &   $1$             &   $-1$ \\
 1+1               &   $1.45\pm0.06$    &   $-1.001\pm0.002$ &   $0.34 \pm0.03$  &   $-0.69\pm0.02$ \\
 2+1               &   $1.55\pm0.35$    &   $-1.00 \pm0.01$  &   $0.17 \pm0.01$  &   $-0.48\pm0.02$ \\
 3+1               &   $1.58\pm0.30$    &   $-1.00 \pm0.02$  &   $0.093\pm0.006$ &   $-0.34\pm0.02$ 
\\ \hline
\end{tabular}
\caption{Asymptotic power-law parameters for $\alpha$ and $\beta$, defined in equations \ref{eq:vacform} 
         and \ref{eq:sigmaform}.}
\label{tab:results}
\end{center}
\end{table}

Extrapolating beyond the range of parameters directly probed in our simulations was unavoidable, since 
running the simulations with large numbers of bubble nucleations is extremely time and memory consuming.  
We ran the simulations for several months on a cluster of 32 1GHz processors.  However, it is clear from 
the analysis of appendix \ref{appendixA} that for any given dimensionality, $\alpha$ asymptotes to 
$\sim(H\tau)^{-1}$, and $\beta$ is consistent with power-law behavior.  Of course it is possible that 
a drastic change in the behavior of the system could take place beyond the range of our simulations, 
but given the simple scaling evident in figures \ref{fig:alpha} and \ref{fig:beta}, this seems rather 
unlikely.

\begin{figure}[t]
\begin{center}
\epsfxsize=0.7\textwidth\epsfbox{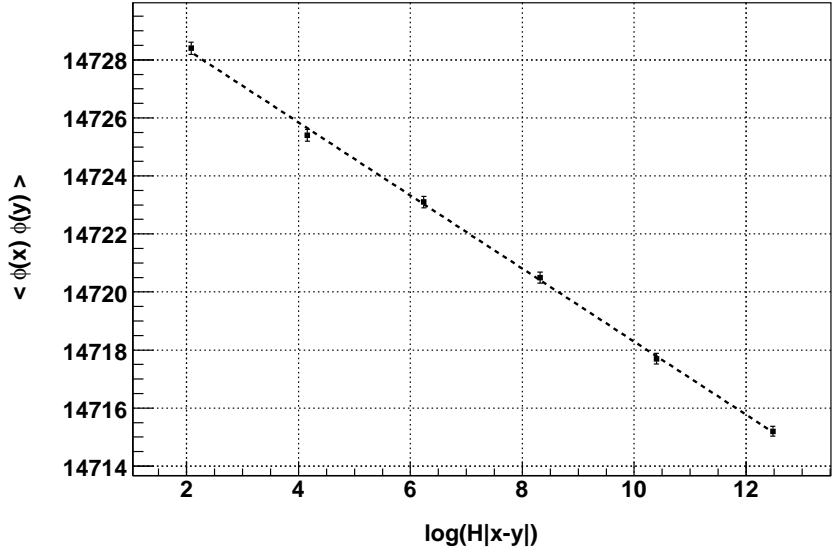}
\caption{Spatial 2-point correlator of $\phi$ as a function of separation.  This was generated by the 
         simulation with 1+1 dimensions, $(H\tau)^{-1} = 6$, $Ht = 16$.  The slope of the best-fit 
         line is $-1.259 \pm  0.022$.  The $\beta$ value for this set of parameters is $1.262 \pm 0.006$.}
\label{fig:correlator}
\end{center}
\end{figure}

Finally, for a few isolated points in parameter space,\footnote{Specifically, in 1+1 we checked 
$(H\tau)^{-1}$ = 6 and 8, $Ht$ = 8 and 16, all four combinations.  In 2+1 we checked $(H\tau)^{-1}$ = 3, 
$Ht$ = 16.} we ran full simulations for the 2-point correlator $\langle\phi(\bfx,t)\,\phi(\bfy,t)\rangle 
- \vev{\phi(t)}^2 $ in order to show agreement with our expression \ref{eq:logform} for how it should be 
determined in terms of $\phi_{RMS}(t)^2$.  A typical result is shown in figure \ref{fig:correlator}.  
Statistical agreement with the expected slope was found for all parameter points that we checked.

\section{Conclusions and Discussion}

We have demonstrated that Chain Inflation, with the assumption of uniformity of tunneling events at 
each step of the chain, requires about $10^4$ vacua per e-folding of inflation.  This is necessary in 
order to produce density perturbations of the size measured by COBE.

Perhaps the primary model in which one might imagine having a uniform sequence of tunneling events is 
the axion model of \cite{Freese2}.  The axion potential takes the form of a tilted cosine, with the 
number of local minima scaling like the number of quark flavors charged under the Peccei Quinn symmetry 
\cite{axion}.  We have thus shown that this model produces unacceptable density perturbations, 
unless the number of fermions in the theory is very large.

While our result for the density perturbations does depend on the assumption of  uniformity of tunneling 
events, we may still put a constraint on more general setups for chain inflation.  It is clear that for 
a fixed total number of transitions, the effect of allowing the tunneling to be non-uniform will be an 
{\it increase} in the size of the perturbations, not a decrease.  Thus in the non-uniform case, $10^4$ 
vacua per e-folding of inflation becomes a {\it lower bound}.

The large number of vacua we have shown to be required by the model is worrisome, as it seems hard to 
imagine such a long sequence of vacua being able to exhibit the properties needed for successful chain 
inflation.  For example, it is necessary that the system never enter a stage of eternal inflation.  
Moreover, since about one e-folding worth of radiation is left over from previous percolations at any 
given time, the energy in the radiation bath will be about $10^4$ times the vacuum energy difference 
between successive steps of the chain.  The effects of such a large quantity of radiation on the 
evolution of the system are thus very important to explore.

There are clearly many open problems in chain inflation, and much room for future work.  Our results 
were obtained through extrapolations of computer simulations, and trying to find an analytic 
understanding of the power law behaviors \ref{eq:vacform} and \ref{eq:sigmaform} seems important, 
although the problem has proven rather difficult.  It is also an open question as to whether or not 
chain inflation could yield observable non-gaussianities in the CMB.  Finally, there is another 
possibility worth considering:  It might seem that the assumption of uniformity of tunneling events 
was necessary in order to obtain a scale invariant spectrum of fluctuations.  On the other hand, 
since we know that many vacuum transitions are required per e-folding of inflation, a given measurement 
of a CMB multipole moment actually averages over the results of a large number of tunneling events.  
It is conceivable that even if the tunneling rates were varying, they might be sufficiently uniform 
``on average'' to produce a scale invariant spectrum.  It would be interesting to explore this 
possibility, perhaps with a string-inspired chain of much more than $10^4$ vacua.

\section*{Acknowledgements} 

We are grateful to Gary Jung and the rest of the Scientific Cluster Support group at Lawrence Berkeley 
National Laboratory.  Our use of their ``minime'' cluster of computers for our simulations was integral 
to the completion of this work.  We would also like to thank Christos Papadimitriou for useful 
discussions.

\newpage

\appendix


\section{Computer Simulation}  \label{appendixA}

Here we shall discuss in more detail the algorithm used in our monte carlo simulations.

The volume of a simulation is a ($D$$-$1)-dimensional cube of {\it physical} side length $2H^{-1}$ 
crossed with an interval of length $t$, the time extent of the simulation.  This is meant to represent 
a $D$-volume in unperturbed de Sitter space in physical coordinates, essentially a region of fixed 
spatial extent about a particular free-falling worldline centered at the origin.  The lightcone of a 
point in this coordinate system can be time-sliced into ($D$$-$1)-spheres of radius 
$H^{-1}\left[e^{H\Delta t} - 1 \right]$, centered about the spatial position of a comoving worldline 
passing through that point.  The set of points with $|\bfx| = H^{-1}$ specifies the causal horizon.  
For any point within this spacetime ``cylinder,'' the past lightcone will be completely contained as well.

We construct this volume to investigate the field fluctuations at the spacetime point $({\mathbf 0},t)$.  
The $D$-volume described above contains the entirety of this point's past lightcone in that time 
interval, which is restricted to lie within the horizon.  Assuming that inflation starts at $t = 0$, 
this $D$-volume then contains all bubble nucleations that will determine the field value, as well as 
{\it their} past lightcones.  An illustration of the shape of $({\mathbf 0},t)$'s past lightcone can
 be seen in figures \ref{fig:dots} and
\ref{fig:dotsAndWalls}.

Since the bubble nucleation density per unit $D$-volume is constant for all tunnelings, we distribute 
nucleation sites in a statistically uniform manner across the volume with no prior knowledge of which 
vacuum is nucleating at each site.  First, we throw the number of sites from a Poisson distribution 
with mean $\Gamma t \left(2H^{-1}\right)^{D-1}$.  We then place these sites at uniformly random, 
uncorrelated positions in the $D$-volume, disregarding those nucleations that fall outside of 
$({\mathbf 0},t)$'s past lightcone.

To calculate the statistics of $\phi({\mathbf 0},t)$, we implement the procedure described in section 
\ref{fluctuations} for a large number (typically tens of thousands) of random nucleation configurations.  
Though following this basic algorithm, the simulation also incorporates a few time-saving tricks.

First, the past lightcone search for each nucleation site, necessary to implement equation \ref{eq:vac2}, 
is cut off at a time $t_{cut}$ earlier.  This is motivated by the fact that successive percolations 
``screen'' each other.  The parent bubble in which a given nucleation will take place will typically 
have been birthed no more than a few $\tau$ in the past.  It is obviously counterproductive to consider 
nucleations that occurred many $\tau$ earlier, as the bubbles generated during those early times will 
have been crowded out by more advanced bubbles.  Consequently, we set $t_{cut}$ to be a few $\tau$, 
typically 3$\tau$ for $D = 4$, and 5$\tau$ for $D = 3$ and $D = 2$.\footnote{In lower dimensions, a 
point in spacetime sees a smaller number of nucleations in its recent past for fixed $t_{cut}$ than a 
point in higher dimensions would, and thus has a greater chance of missing the nucleation(s) with 
maximum $\phi$.}  We find that these values yield statistical results that are stable under $t_{cut}$ 
increases.

The second trick is to divide ($D$$-$1)-space into numbered cubical ``boxes'' in order to organize the 
nucleation sites.  The boxes are ${\cal O}(t_{cut})$ along each side, so that we can find all sites in 
a past lightcone out to $t_{cut}$ by searching just within boxes in the immediate vicinity.  Distant 
sites are thus automatically ignored.  More rigorously, we calculate the spatial extent of the past 
lightcone $t_{cut}$ back, and search over all boxes that might contain it.  Since this procedure 
captures the entire past lightcone out to $t_{cut}$, it does {\it not} constitute an additional 
approximation.

We also perform a handful of modified simulations in order to calculate the 2-point correlator $\langle 
\phi(A) \, \phi(B) \rangle$ for spacetime points $A$ and $B$ outside of each other's respective 
horizons.  These follow largely the same procedure, with simulations for $A$ and $B$ run in parallel.  
Points that actually share history will have common nucleations in their past lightcones.  Consequently, 
some of the nucleation sites in $A$'s and $B$'s simulation volumes are actually ``the same,'' and we 
correlate their placement accordingly.


\section{Extrapolation Method} \label{appendixB}

Here we present the analysis techniques used to extract the results for table \ref{tab:results} in 
section \ref{results}.

We ran the monte carlo simulation in $D = 2,3,4$, with nucleation rates $(H\tau)^{-1} = 3,6,12,\ldots$, 
and for time durations $Ht = 1,2,4,8,\ldots$, as far as our processor and memory resources would allow 
within a reasonable time.  The set of measurements for $\vev{\phi(t)}$ and $\phi_{RMS}(t)^2$ appear in 
figures \ref{fig:meanVsT2D} through \ref{fig:sigmaVsT4D}.  We fit these to obtain the slopes 
$\alpha(H\tau)$ and $\beta(H\tau)$, presented in figures \ref{fig:alpha} and \ref{fig:beta}.

\begin{figure}[p]
\begin{center}
\epsfxsize=0.7\textwidth\epsfbox{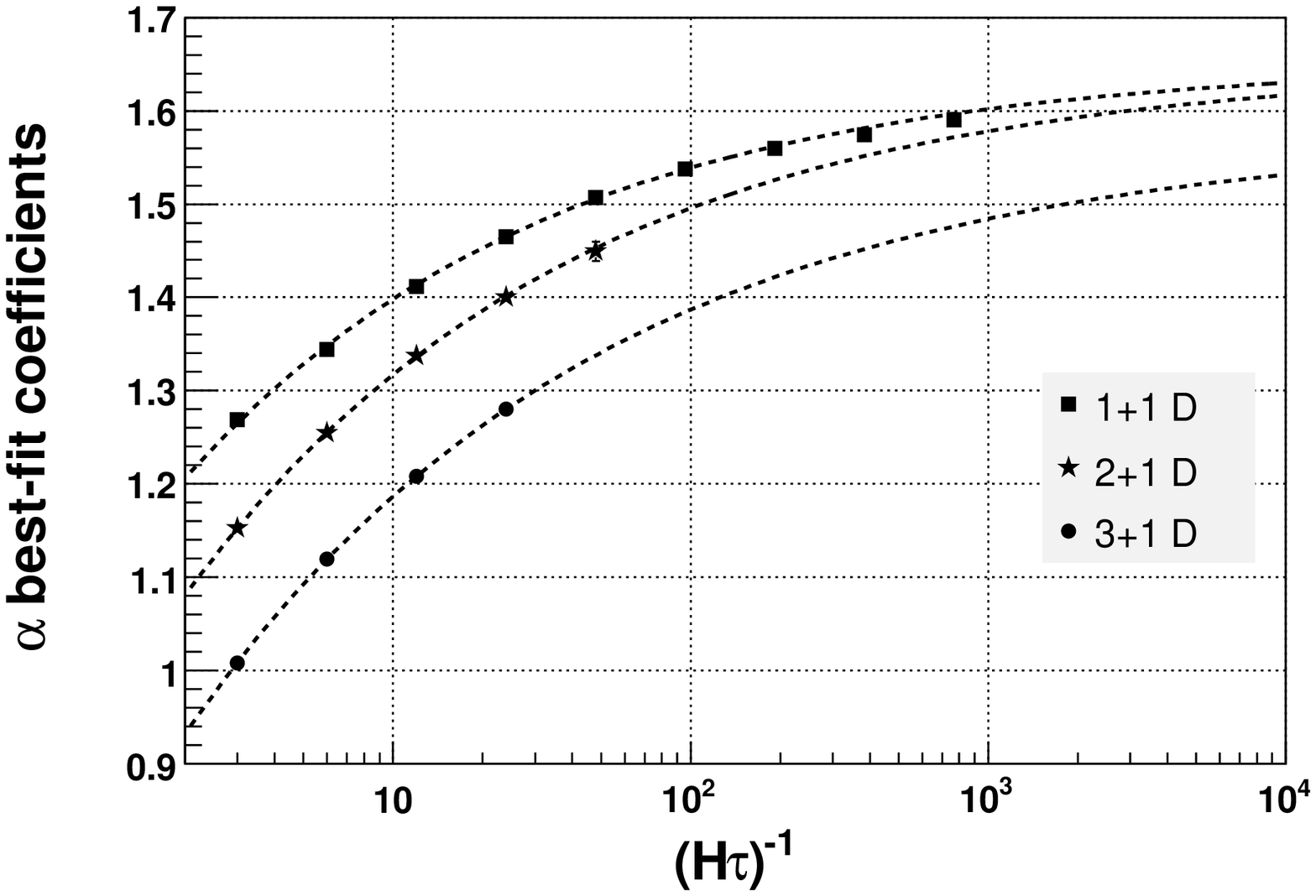}
\caption{Best-fit power-law coefficients for $\alpha$ using pairs of points $(H\tau)^{-1}$ and 
         $2(H\tau)^{-1}$.   For clarity, the points for 1+1 dimensions and 2+1 dimensions have been
         offset vertically by 0.2 and 0.1, respectively.  The extrapolations (dashed lines) are 
         described in the text.}
\label{fig:alphaCoeffs}
\vspace{0.5in}
\epsfxsize=0.7\textwidth\epsfbox{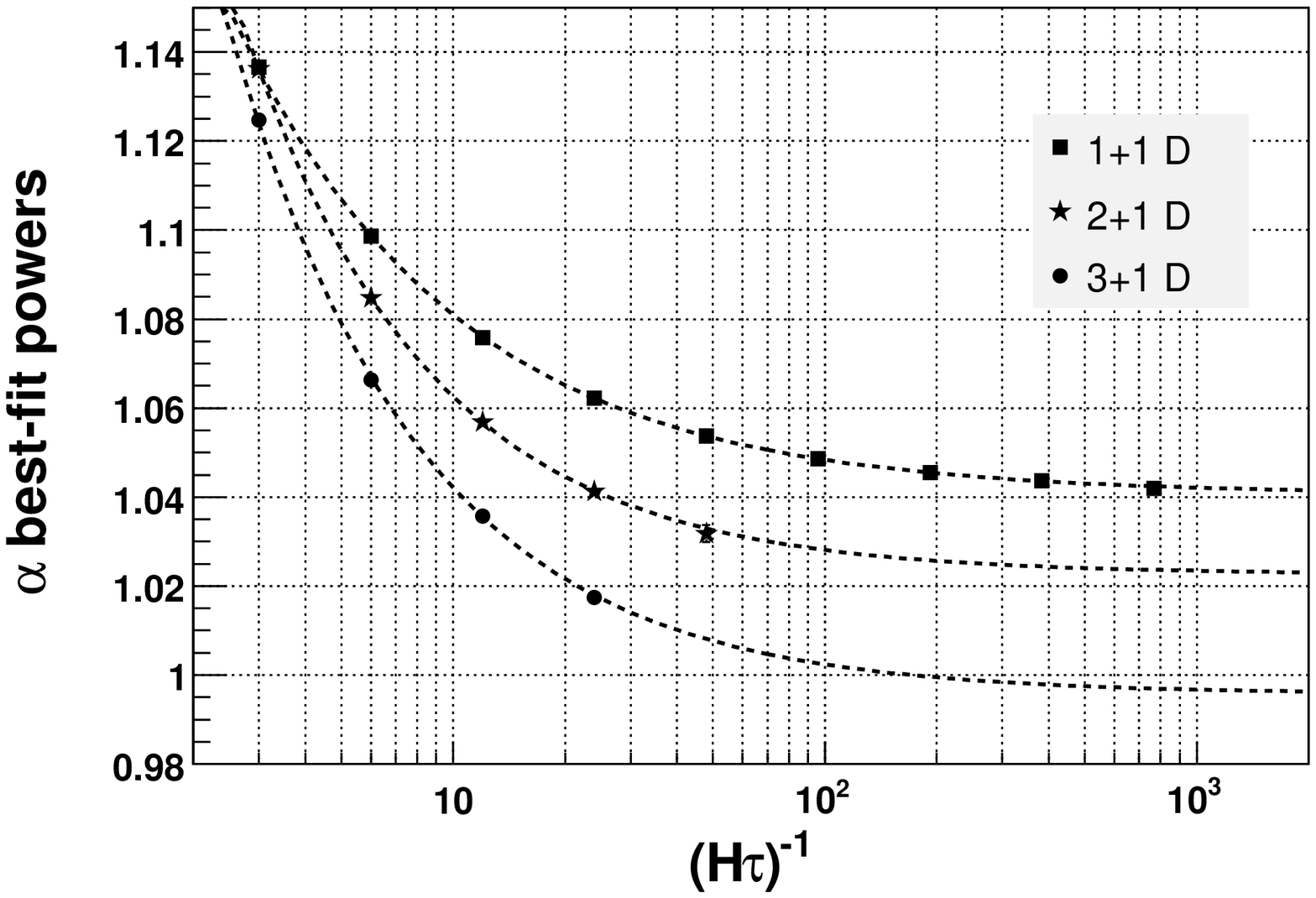}
\caption{Best-fit power-law powers for $\alpha$ using pairs of points $(H\tau)^{-1}$ and 
         $2(H\tau)^{-1}$.  For clarity, the points for 1+1 dimensions and 2+1 dimensions have been 
         offset vertically by 0.04 and 0.02, respectively.  The extrapolations (dashed lines) are 
         described in the text.}
\label{fig:alphaPowers}
\end{center}
\end{figure}

\begin{figure}[p]
\begin{center}
\epsfxsize=0.7\textwidth\epsfbox{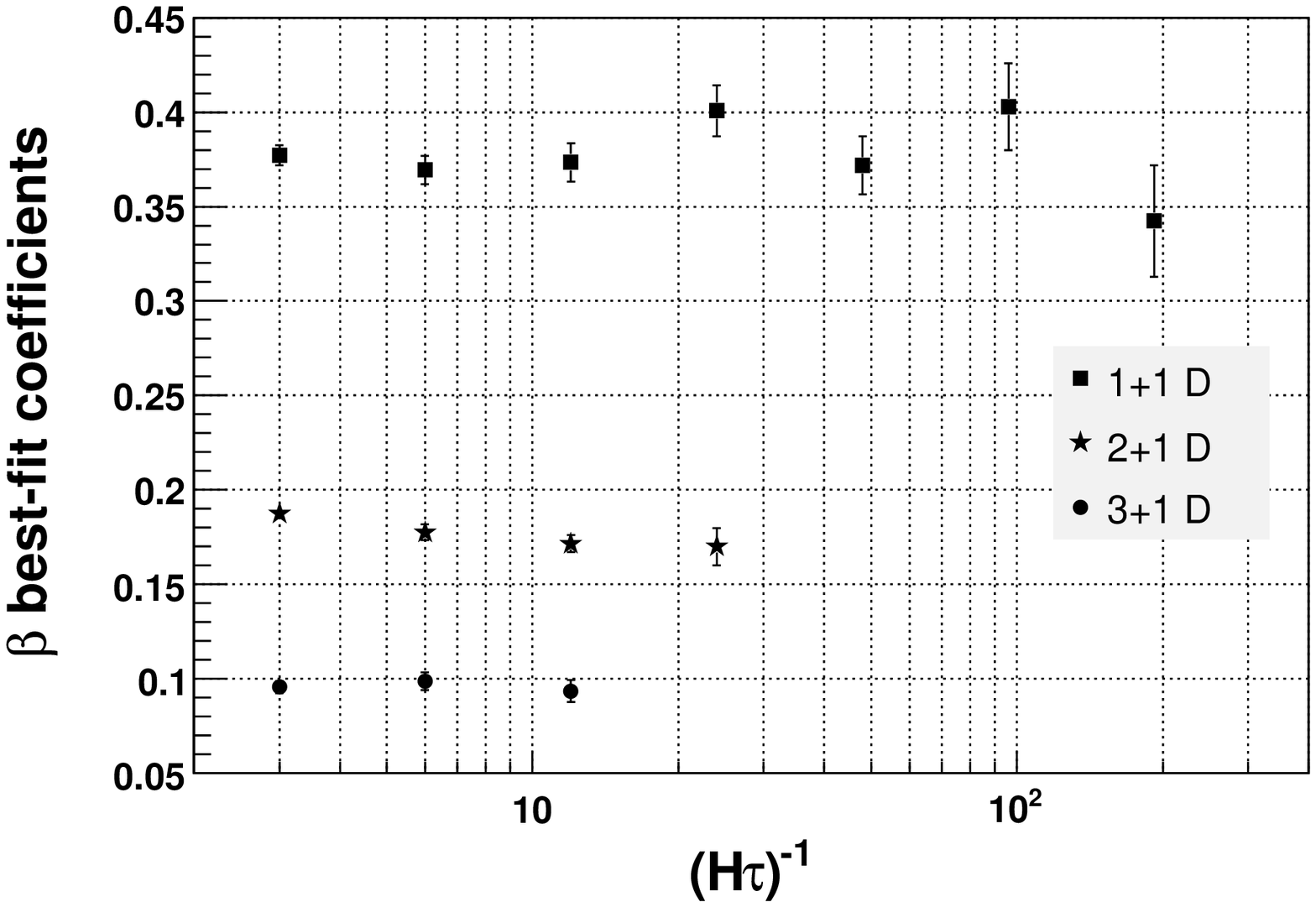}
\caption{Best-fit power-law coefficients for $\beta$ using pairs of points $(H\tau)^{-1}$ and 
         $2(H\tau)^{-1}$.}
\label{fig:betaCoeffs}
\vspace{0.5in}
\epsfxsize=0.7\textwidth\epsfbox{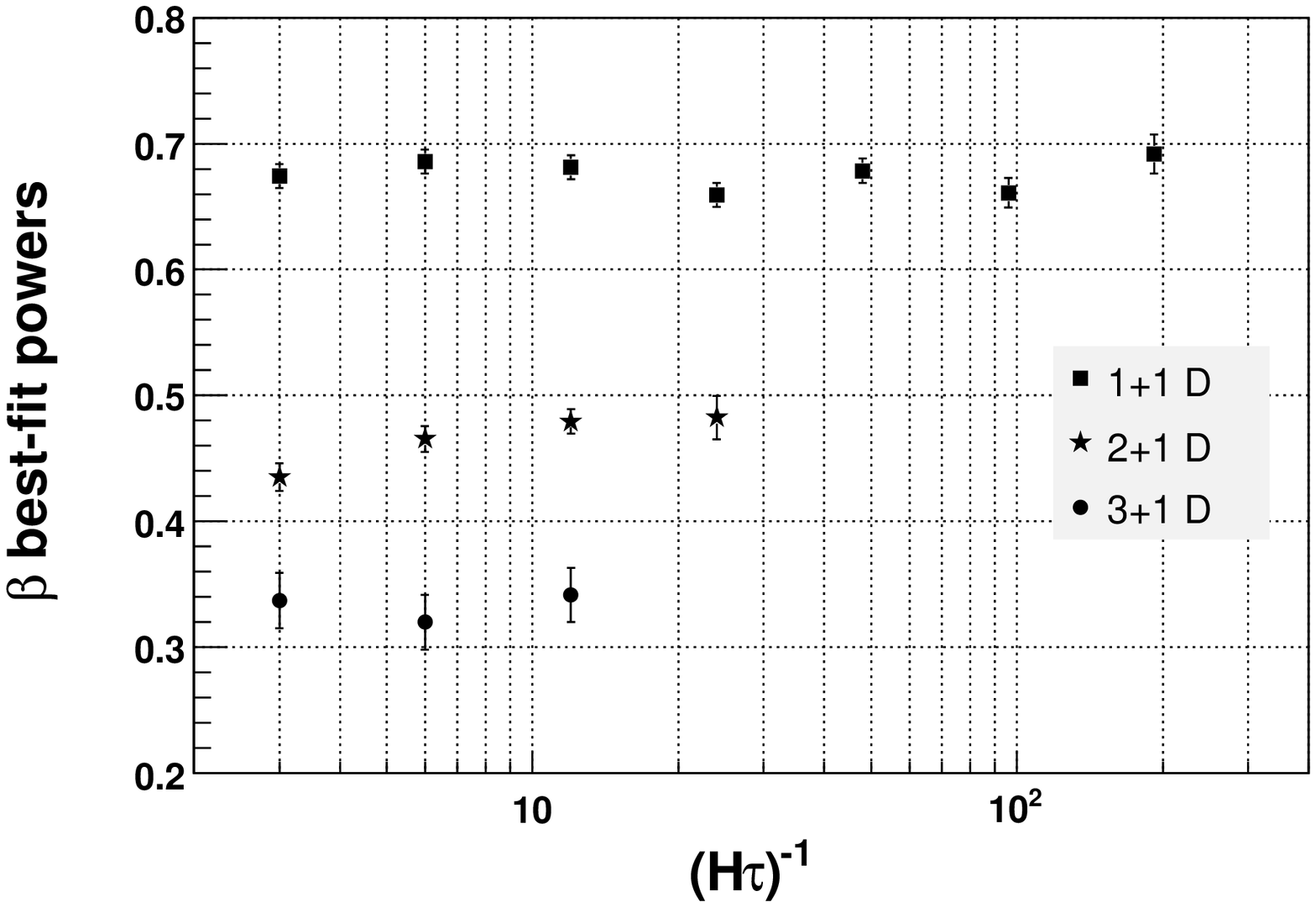}
\caption{Best-fit power-law powers for $\beta$ using pairs of points $(H\tau)^{-1}$ and 
         $2(H\tau)^{-1}$.}
\label{fig:betaPowers}
\end{center}
\end{figure}

In $D = 4$, the largest $(H\tau)^{-1}$ for which our statistics allowed {\it both} $\alpha$ and $\beta$ 
to be precisely calculated was $24$.  Equation \ref{eq:result} indicates that the resulting values are 
far from reproducing the correct scale of density perturbations.  Consequently, we are forced to perform 
an extrapolation to much larger values of $(H\tau)^{-1}$.  We assume that $\alpha$ and $\beta$ asymptote 
to the power law forms of equations \ref{eq:vacform} and \ref{eq:sigmaform}.  In order to estimate the 
asymptotic power laws, we first fit ascending pairs of $(H\tau)^{-1}$ points to power laws, and then 
investigate the convergence of the coefficients and powers returned by the fits.  The results of these 
pairwise fits are illustrated in Figures \ref{fig:alphaCoeffs} through \ref{fig:betaPowers}.  Note that 
for $\alpha$, points with different $D$'s are vertically offset for clarity of presentation.  In reality, 
they essentially lie on top of each other.

For $\alpha$, the statistical errors are small and the convergence behavior is clear.  However, for the 
physically relevant $D = 4$ case, we were not able to go far enough in $(H\tau)^{-1}$ to explicitly see 
the convergence as well as we can in lower dimensions.  In order to perform the extrapolation, we fit 
the power-law parameters in all 3 $D$'s with functions of the form $p_1 + p_2 (H\tau)^{p_3}$, where the 
$p_i$ are the fit parameters.  These are seen to be visually good matches in the figures, though note 
that we do {\it not} claim that they furnish accurate models of the convergence behavior.  (The 
$\chi^2/NDF$ values for these fits range from 1 to 100.)  The best-fit asymptotic values $p_1$ 
constitute our central estimates.  We then define the systematic error on each of these as the difference 
between the extrapolation estimate and the power-law parameter measured for the last pair.  Our estimate 
for the $D = 4$ asymptotic coefficient and power are thus $1.58\pm0.30$ and $-1.00\pm0.02$, 
respectively.  Interestingly, the results from all dimensions taken together suggest a universal 
asymptotic form $\alpha(H\tau) \simeq (1.5)(H\tau)^{-1}$, though we do not assume this in our analysis.  
In this limit, $\vev{\phi(t)}$ is independent of $H$.  In any event, the actual dependence on $H$ is 
seen to be very weak.

For $\beta$, the statistical errors are too large to resolve any details of the hypothesized convergence 
behavior.  All pairs of points yield consistent coefficients and powers for the power-law 
fits.\footnote{The $D = 3$ points do display some slope by eye, and a fit to a flat line indicates 
that convergence across the entire sampled range of $(H\tau)^{-1}$ is only about 1\% probable.  
However, consistency with a flat line does emerge past the first point.}  We therefore 
conservatively define the central estimates and their errors as the parameter values and errors obtained 
from the power-law fit of the last pair.  For $D = 4$, the estimated asymptotic coefficient and power 
are $0.093\pm0.006$ and $-0.34\pm0.02$, respectively.

Another source of potential systematic error comes from the $t_{cut}$ cutoff on past lightcone searches, 
described in appendix \ref{appendixA}.  We examined the numerical stability of our $\alpha$ and $\beta$ 
values with respect to variations in $t_{cut}$ in all $D$ for several values of $\tau$, and found 
convergence to within the statistical errors.  Thus, we do not associate an error with this variable.  
If it is there, it is subdominant.

The remainder of our estimated asymptotic power-law forms for $\alpha$ and $\beta$ are summarized in 
table \ref{tab:results}.


\newpage


\begin{thebibliography}{99}

\bibitem{Guth}
%
A. ~Guth,
  ``The Inflationary Universe: A Possible Solution to the Horizon and Flatness Problems,''
  Phys.\ Rev.\ D {\bf 23}, 347-356 (1980); \\
%
\bibitem{Coleman}
S. ~Coleman,
  ``The Fate of the False Vacuum. 1. Semiclassical Theory,''
   Phys.\ Rev.\ D {\bf 15}, 2929-2936 (1977); \\
%
\bibitem{GW}
A. ~Guth, E. ~Weinberg,
   ``Could the Universe Have Recovered from a Slow First Order Phase Transition,''
   Nucl.\ Phys.\ B {\bf 212}, 321 (1983); \\
%
\bibitem{slowroll1}
A. ~Linde,
  ``A New Inflationary Universe Scenario:  A Possible Solution of the Horizon, Flatness, Homogeneity, 
    Isotropy and Primordial Monopole Problems,''
   Phys.\ Lett.\ B {\bf 108}, 389-393 (1982); \\
%
\bibitem{slowroll2}
A. ~Albrecht, P. ~Steinhardt,
  ``Cosmology for Grand Unified Theories with Radiatively Induced Symmetry Breaking,''
   Phys.\ Rev.\ Lett.\ {\bf 48}, 1220-1223 (1982); \\
%
\bibitem{Freese1}
K. ~Freese, D. ~Spolyar,  
 ``Chain Inflation: Bubble, Bubble, Toil and Trouble,''
  JCAP.\ {\bf 0507}, 007 (2005)
  [hep-ph/0412045]; \\
%
\bibitem{Freese2}
K. ~Freese, J. T. ~Liu, D. ~Spolyar,  
  ``Inflating with the QCD Axion,''
  Phys.\ Rev.\ D {\bf 72}, 123521 (2005); \\
%
\bibitem{normalization}
E. ~Bunn, A. ~Liddle, M. ~White, 
  ``Four Year COBE Normalization of Inflationary Cosmologies,''
  Phys.\ Rev.\ D {\bf 54}, 5917-5921 (1996)
  [astro-ph/9607038]; \\
%
\bibitem{Hawking}
S. ~Hawking, 
  ``The Development of Irregularities in a Single Bubble Inflationary Universe,''
   Phys.\ Lett.\ B {\bf 115}, 295 (1982); \\ 

%
\bibitem{Starobinsky}
A. ~Starobinsky,
  ``Dynamics of Phase Transition in the New Inflationary Universe Scenario and Generation of 
    Perturbations,''
   Phys.\ Lett.\ B {\bf 117}, 175-178 (1982); \\
%
\bibitem{GuthPi}
A. ~Guth, S. ~Pi,
  ``Fluctuations in the New Inflationary Universe,''
   Phys.\ Rev.\ Lett.\ {\bf 49}, 1110-1113 (1982); \\
%
\bibitem{LR}
D. ~Lyth, A. ~Riotto,
  ``Particle Physics Models of Inflation and the Cosmological Density Perturbation,''
  Phys.\ Rept.\ {\bf 314}, 1-146 (1999)
  [hep-ph/9807278]; \\
%
\bibitem{axion}
P. ~Sikivie, 
   ``Of Axions, Domain Walls and the Early Universe,''
    Phys.\ Rev.\ Lett.\ {\bf 48}, 1156 (1982); \\

\end{thebibliography}
\end{document}